\documentclass[prb,twocolumn,showpacs,preprintnumbers,amsmath,amssymb]{revtex4}
\usepackage{graphicx}
\usepackage{amsmath}\usepackage{amsbsy}
\usepackage{dcolumn}
\usepackage{bm}

\begin{document}

\def\iom{\imath\omega}
\def\iOm{\imath\Omega}
\def\iomp{\imath\omega^{\prime}}
\def\omp{\omega^{\prime}}
\def\id{\imath\delta}
\def\sp{s^{\prime}}
\def\taup{\tau^{\prime}}
\def\dt{\partial\over\partial\tau}
\def\dtp{\partial\over\partial\tau^{\prime}}
\def\tpi{2\pi\imath}
\def\I{{\rm I}}
\def\z{{\kappa}}
\def\Ss{\Lambda}
\def\a{\alpha}
\def\bk{{\bf k}}
\def\bq{{\bf q}}
\def\ek{\varepsilon_{\bf k}}
\def\Jq{J_{\bf q}}
\def\uk{_{\bf k}}
\def\uq{_{\bf q}}
\def\vk{{\bf k}}
\def\vq{{\bf q}}
\def\Tr{{\rm Tr}}
\def\hph{{\bf\Phi}}
\def\D{{\cal D}}
\def\kl{l}
\def\L{{\cal L}}

\newcommand{\w}{\omega}
\newcommand{\sign}{\text{sign}}
\newcommand{\s}{\sigma}
\newcommand{\RE}{\text{Re}}
\newcommand{\IM}{\text{Im}}
\newcommand{\TK}{T_K}
\newcommand{\g}{\gamma}
\newcommand{\G}{\Gamma}
\newcommand{\Nf}{N_F}
\newcommand{\pf}{PF}
\renewcommand{\vec}[1]{{\mathbf #1}}


\title{
Pseudogaps in the $t$-$J$ model: Extended DMFT study}

\author{K. Haule$^{1,4}$, A. Rosch$^{2}$, J. Kroha$^{3}$ and P. W\"olfle$^{2}$}

\affiliation{$^{1}$ Department of Physics and Astronomy,
Rutgers  University,
136 Frelinghuysen Road,
Piscataway NJ 08854-8019, USA\\
$^{2}$Institut f\"ur Theorie 
der Kondensierten Materie, Universit\"at Karlsruhe, 
D-76128 Karlsruhe\\
$^{3}$ Physikalisches Institut,
Universit\"at Bonn,
Nussallee 12,
53115 Bonn\\
$^{4}$ J. Stefan Institute, 1000 Ljubljana, Slovenia}

\date{\today}

\begin{abstract}
We investigate the highly incoherent regime of hole-doped 2d
Mott-Hubbard insulators at moderately small doping $\delta$ and temperatures
$\gtrsim 0.1J$, where $J$ is the exchange coupling.
Within an extended dynamical mean-field theory of the $t$-$J$ model and
a generalized non-crossing approximation we calculate the
single-particle 
spectral function,   the dynamical susceptibility, and
thermodynamic and transport quantities.  Short-ranged
antiferromagnetic fluctuations lead to strongly incoherent
single-particle 
dynamics,  large entropy and large electrical resistivity.
At low doping a pseudogap is found to open up both in the
single-particle 
and the spin excitation spectra leading to a decrease in
entropy and resistivity.  The Hall coefficient changes sign to
positive values  upon lowering   the doping level and increases
inversely proportional to $\delta$.   
\end{abstract}

                                 

\maketitle

\section{Introduction}

Strongly interacting Fermi systems on a two-dimensional lattice have
been a focus of interest ever since high temperature superconductor
materials were discovered in 1986.  The unusual properties of these
materials, in particular in the normal conducting phase, have led to
the hypothesis that the usual Landau Fermi liquid theory may not be
applicable in this case \cite{A}.  A widely accepted view holds that
these systems may be considered as hole-doped Mott-Hubbard
insulators. The correlations present in a Mott-Hubbard insulator are
characterized by strongly suppressed charge fluctuations (due to the
constrained hopping of the holes resulting from the strong on-site
Coulomb repulsion U) and enhanced quantum spin fluctuations governed
by the antiferromagnetic nearest-neighbor spin exchange interaction.
The interplay of the motion of holes with the antiferromagnetically
correlated spin background is the central problem of hole-doped
Mott-Hubbard insulators.  Despite an extraordinary effort by many
theorists and a correspondingly large number of papers we believe it
to be fair to say that a thorough understanding of this problem is
still lacking.

The ground state of the Mott-insulating state of electrons on a square
lattice at half-filling is expected to be antiferromagnetically
ordered \cite{SHN,BBGW}.  Doping with holes leads to a rapid
destruction of long-range order, at a critical concentration
$\delta_c$ of a few percent doping.  For larger dopings there is
evidence for strong antiferromagnetic spin fluctuations of relatively
short range.

In this paper we undertake to explore the consequences of strong
incoherent and local (i.e. nearest neighbor) spin fluctuations on the
dynamics of charge carriers, and on the thermodynamics of the systems.
We also investigate how the single-particle properties feed back into
the spin dynamics.  Our approach is focused on the temperature regime
of $0.1 J \lesssim T \lesssim t$ ($J = 0.3t$ in the cuprates) where
$J$ is the exchange constant and $t$ is the nearest neighbor hopping
amplitude.  In this regime we expect strong quantum and thermal
fluctuations driven by competing interactions to decohere the fermionic
excitations.  This temperature regime is bounded from below by
possible antiferromagnetic, superconducting or other ordered states.
The incoherent regime is confined to small doping levels $\delta_c
\lesssim \delta \lesssim 0.3$, and crosses over into a Fermi liquid
state at $\delta\gtrsim 0.3$.

A minimal model encompassing the physics described above is the $t$-$J$
model.  It is well known that the interplay of hopping and the local
correlations induced by the on-site Coulomb interaction may be
captured in Dynamical Mean-Field Theory, in which the lattice model is
mapped onto a quantum impurity coupled to a fermionic bath in a
self-consistent fashion \cite{MV,GGKR}.  In the same spirit the
nearest neighbor exchange interaction of a given spin to its
neighboring spins may be approximated by a dynamically fluctuating
bosonic field, to be determined self-consistently \cite{SS,K}.  In
this way the two principal processes, constrained hopping and spin
exchange interaction, may be fully incorporated on the same footing,
on the level of short-range correlations.

As reviewed in Section \ref{section.EDMFT}, the Extended Dynamical
Mean Field Theory (EDMFT) for the two-dimensional $t$-$J$ model is
obtained by approximating the single-particle self-energy
$\Sigma_{\vec k}(\omega)$ and the two-particle self-energy $M_{\vec
  q}(\omega)$ by momentum independent functions.  $\Sigma(\omega)$ and
$M(\omega)$ are obtained by equating the local (i.e. the momentum
integrated) single-particle Green's function and spin susceptibility,
respectively, with the corresponding quantities of an extended
Anderson impurity model, featuring a fermionic and a bosonic bath to
be determined self-consistently.  For the bare hopping integrals and
exchange couplings we use a nearest-neighbor tight-binding model on
the square lattice.  The local approximation is  better the higher
the spatial dimension $d$ and becomes exact for $d\rightarrow \infty$,
provided the hopping amplitude $t$ and the exchange coupling $J$ are
scaled as $t/\sqrt{d}$ and $J/\sqrt{d}$.  This scaling is possible in
the paramagnetic regime.  Most of the methods employed for the
solution of the Anderson impurity or Kondo problem do not work here.
We use self-consistent perturbation theory in the form of conserving
approximations \cite{CKW,KW}, and the exact projection onto the
Hilbert space without double occupancy (limit $U\rightarrow \infty$).
We are interested in describing the highly incoherent regime at small
doping levels and not too low temperatures, where the spectral
functions are broad and relatively featureless.  In this regime we
expect vertex corrections and higher order processes in general, to
change the characteristic parameters like maximum values, peak widths
and 
gap widths of the dynamic quantities by correction terms of order
unity, but we do not expect that these contributions lead to more
coherence or new collective behavior.  In this spirit we approximate
all self-energies by their lowest order self-consistent perturbation
theory expressions (in the hopping parameter and exchange coupling).
The resulting theory, presented in Section \ref{s.NCA}, is an
extension of the Non-Crossing Approximation (NCA) \cite{B} including
the bosonic bath.

The results of this approximation scheme for the $t$-$J$ model are
presented in Section \ref{s.results}.  It turns out that nearest-neighbor spin
fluctuations are sufficient to create a pseudogap in the
single-particle spectrum and in the spin excitation spectrum at $q$-vectors
away from $(\pi,\pi)$, for small dopings $\delta\lesssim 0.1$, similar
to what is seen in ARPES experiments \cite{I} and in the magnetic
properties \cite{TM,TL}.  The pseudogap scales with $J$. There are
several indications that Fermi liquid behavior is violated for
$\delta\lesssim 0.2$. Most noteworthy, the effective chemical
potential is found to move from the center of the band up to the band
edge, as the doping is decreased to small values.  As $\delta$ grows
beyond $0.25$, however, Fermi liquid behavior appears to be restored.
The entropy turns out to be large in the range $0.1\lesssim
\delta\lesssim 0.2$ and is reduced on both sides of this interval by
the pseudogap and the incipient Fermi liquid behavior, respectively.
The resistivity is dominated by strong incoherent scattering, and the
Hall coefficient is found to be hole-like, $\propto 1/\delta$, for
small $\delta$, again resembling the observed behavior \cite{CL}.
Some of the results have been reported in [\onlinecite{HRKW}].

Results similar to ours have been found in two recent works using DMFT
for a clusters of sites within the Hubbard model.  Maier {\it et al.}\cite{Maier}  applied the dynamical
cluster approximation (DCA) for various cluster sizes up to 64 sites to
the Hubbard model in the intermediate coupling regime ($U\sim
\text{bandwidth}$).  The DCA equations were solved with Quantum Monte
Carlo techniques down to room temperature.
The authors of Ref.~[\onlinecite{Maier}] identified signals for
non-Fermi liquid behavior at low doping $\delta \lesssim 0.1$ and
found a large residual scattering rate and a pronouced pseudogap at
low doping.  In Ref. [\onlinecite{Phillips}], Stanescu and Phillips studied
the Hubbard model at intermediate coupling within a two-site cluster
approach using the Non-Crossing Approximation as a quantum impurity
solver at not too low temperatures.  It is again found that
Luttinger's theorem appears to be violated for low doping in a regime
where a pseudogap opens.

Despite the similarity of the numerical results, quite different
explanations for the observed pseudogap physics have been suggested,
ranging from short-range spin correlations, spin-charge separation and
resonance valence bond (RVB) physics\cite{Maier} to effects of the
upper Hubbard band and current-correlations involving three
neighboring sites\cite{Phillips}. By construction our approximation scheme is 
 not able to describe such intersite correlations or RVB
singlets and does not include the upper Hubbard band: nevertheless
the overall results are qualitatively very similar. We take this as a
strong indication that neither short-range magnetic or current
correlations nor RVB physics is the underlying reason but argue that
there is another generic mechanism for pseudogap formation: The
strongly incoherent dynamics captured in our scheme as well as those
of Refs. [\onlinecite{Maier,Phillips}] appears to be the dominant
feature of the Hubbard model as well as the $t$-$J$ model in the low
energy sector $(0.1 J \lesssim \omega \lesssim t)$ for small doping.
Therefore pseudogap formation seems to be a generic property of any
strongly incoherent Fermi system close to a Mott insulator. In other
words, the existence of a pseudogap neither requires slowly
fluctuating, finite-ranged ordered domains (antiferromagnetic,
superconducting) \cite{R} nor a local resonance state.

\section{Extended Dynamical Mean-Field Theory of the $t$-$J$ Model}
\label{section.EDMFT}

The standard model embodying the physics of the hole-doped
Mott-Hubbard insulator is the $t$-$J$ model, defined by the Hamiltonian
\begin{equation}
H = \sum_{i,j}t_{ij} \tilde c_{i\sigma}^+ \tilde c_{j\sigma} +
\frac{1}{2} \sum_{i,j} J_{ij}\vec S_i \vec S_j
\label{1}
\end{equation}
where $\vec S_i = \frac{1}{2} \sum_{\sigma,\sigma{'}} \tilde
c_{i\sigma}^+ \vec\tau_{\sigma\sigma{'}}\tilde c_{i\sigma{'}}$ is the
spin operator at lattice site $i$, $\vec \tau$ denotes the vector of
Pauli matrices and $t_{ij}(J_{ij})$ are the hopping amplitudes
(exchange interaction) connecting sites $i$ and $j$. For the numerical
evaluation to be discussed later we will use a tight-binding model on
a two-dimensional square lattice, $t_{ij} = - t\delta_{i,i+\tau}$,
$J_{ij} = J\delta_{i,i+\tau}$, where $\tau$ labels nearest neighbor
sites. The operator $\tilde c_{i\sigma}^+ (\tilde c_{i\sigma})$
creates (annihilates) an electron at site $i$ with spin projection
$\sigma$ at a singly occupied lattice site.  In terms of usual
electron operators $c_{i\sigma}^+ (c_{i\sigma})$ one has $\tilde
c_{i\sigma}^+ = c_{i\sigma}^+ (1-n_{i,-\sigma})$, where $n_{i\sigma} =
c_{i\sigma}^+ c_{i\sigma}$ is the occupation number operator.  In this
way occupation of lattice sites by two electrons with spins $\uparrow$
and $\downarrow$ is avoided, which would cost the large Hubbard energy
$U$.  We will be interested in electron densities close to
half-filling of the band, such that $\langle \sum_\sigma
n_{i\sigma}\rangle = n = 1 - \delta$, where $\delta \ll 1$ is the
doping concentration of holes.

Whereas at exactly half-filling, when $H$ reduces to the Heisenberg
model, the ground state has antiferromagnetic long-range order, we
anticipate that this will not be the case for sufficiently large
doping $\delta > \delta_c$ (in experiment $\delta_c \simeq 0.03$).  In
this regime it is reasonable to assume the antiferromagnetic
correlations in the system to be short-ranged.  We assume furthermore
that additional forms of long-range order (such as superconductivity) that may
be possible ground states of the $t$-$J$ model are confined to a
lower temperature regime, such that the corresponding fluctuations are
sub-dominant at elevated temperatures.  Consequently, one expects an
extended high-temperature regime where short-ranged spin fluctuations
lead to a highly incoherent metallic state, as observed in
high-temperature superconductors, with anomalous transport properties
(large, non-Fermi liquid type electrical resistivity, hole-like Hall
constant), large entropy, broad ``quasiparticle'' peaks in
photoemission, etc.  It is our aim to investigate this regime within
an approximation scheme which neglects most of the longer range
spatial correlations, but keeps the dominant short range spin
correlations.

The single-particle dynamics and the two-particle dynamics of the
model are described by the Green's function
\begin{eqnarray}
G_{\vec k,\sigma}(i\omega) &=& - \int_0^\beta d\tau e^{i\omega\tau}
\left\langle T_\tau \tilde c_{\vec k\sigma}(\tau)\tilde c_{\vec
k\sigma}(0)\right\rangle\nonumber \\
&=& \frac{1}{i\omega +\mu - \epsilon_{\vec k} - \Sigma_{\vec k,\sigma}(i\omega)}
\label{2}
\end{eqnarray} 
and by the spin susceptibility
\begin{eqnarray}
\chi_{\vec q,\alpha}(i\Omega) &=& \int_0^\beta d\tau e^{i\Omega\tau}
\left\langle T_\tau S_{-\vec q,\alpha}(\tau)S_{\vec
q,\alpha}(0)\right\rangle\nonumber \\ &=& \frac{1}{J_{\vec q} +
M_{\vec q,\alpha}(i\omega)}.
\label{3}
\end{eqnarray}
Here $\beta$ is the inverse temperature $T$ (we employ units with $k_B
= \hbar = 1$), $\omega$ and $\Omega$ are fermionic and bosonic
Matsubara frequencies, and $\epsilon_{\vec k}$ and $J_{\vec q}$ are the
lattice Fourier transforms of the hopping amplitudes $t_{ij}$ and the
exchange couplings $J_{ij}$, respectively.  While the self-energies
$\Sigma_{\vec k}(i\omega)$ and $M_{\vec q}(i\omega)$ are momentum
dependent in general, the observation that the fluctuations in the
system are short-ranged in the regime we are interested in suggests
that a ``local'' approximation, neglecting the momentum dependence of
$\Sigma$ and $M$ altogether, may be a good starting point.  We
therefore employ in this paper the main approximation
\begin{equation}
\Sigma_{\vec k} (\omega) \simeq \Sigma (\omega)
\label{4}
\end{equation}
and
\begin{equation}
M_{\vec q}(\omega) \simeq M(\omega)\;\;,
\label{5}
\end{equation}
thus capturing the effect of local fluctuations in time, which
we expect to be important in the presence of strong inelastic
scattering.

The momentum independence of $\Sigma$ and $M$ allows us  to map the
lattice problem onto an Anderson impurity problem where the host
medium has to be determined self-consistently.  Considering first the
single-particle properties, i.e. $\Sigma(\omega)$,  the
corresponding Dynamic Mean-Field Theory (DMFT) has been widely used to
calculate properties of the Hubbard model and periodic Anderson model
\cite{MV,GGKR}. One maps the problem onto an Anderson impurity
embedded in a fermionic bath.  Applied to the $t$-$J$ model it amounts to
treating the exchange interaction in mean-field theory.  This is not
sufficient to allow us to maintain the balance between dynamical hopping
processes and spin fluctuations, which is at the heart of the $t$-$J$
model.  We therefore follow Refs. [\onlinecite{SS,K}] and extend the dynamical
mean-field idea for the paramagnetic phase by introducing a
fluctuating magnetic field coupling to the local spin as representing
an additional class of degrees of freedom of the medium.  This type of
approximation, termed ``extended DMFT'' (EDMFT), has been applied to
the Kondo lattice model \cite{SS} and the extended Hubbard model
\cite{K}. It is important to note that EDMFT becomes exact in the
limit of infinite dimensions $d\rightarrow \infty$, provided $t$ and
$J$ are scaled as $t/\sqrt{d}$ and $J/\sqrt{d}$, respectively.  We
shall use this property in deriving the EDMFT equations (see Appendix
\ref{EDMFT_derivation}). We will, however, regard EDMFT as an
approximation applied in finite dimensions, and as such will use the
tight-binding expressions for $\epsilon_{\vec k}$ and $J_{\vec q}$
valid in $d=2$.

To summarize, the EDMFT is probably best visualized by considering a single-site,
the ``impurity'', and its coupling to the surrounding ``medium''.
There are two types of coupling processes, as is evident from the
Hamiltonian:
\vskip .2cm
\noindent 
(i) hopping to and from the ``impurity'' into the medium, as in the
Anderson impurity model (in the limit of infinite $U$, as a
consequence of the no double occupancy constraint).  The
medium is modeled by a non-interacting fermion system (the
``conduction electrons''), whose local density of states has to be
determined self-consistently.
\vskip .2cm
\noindent 
(ii) exchange coupling of the local spin at the ``impurity'' site to
the spins of the medium.  In the limit $d\rightarrow \infty$ the two
components of the medium, fermions (see above) and spin fluctuations
are completely decoupled. We do not expect that this approximation
holds in 2d for low temperatures.  But in the  regime considered in this
paper, where electrons are highly incoherent, we believe that such a
modeling is appropriate.    The spin fluctuations of the medium is 
described by a (vector) bosonic bath, whose spectrum again has to be
determined self-consistently.  
\vskip .2cm
\noindent 
In this way one is led to a generalized quantum impurity model with
Hamiltonian:
\begin{eqnarray}
H_{\text{EDMFT}} = \sum_{k\sigma} E_k c_{k\sigma}^{\dagger} c_{k\sigma} + V
\sum_{k\sigma} (c_{k\sigma}^{\dagger} \tilde d_\sigma + h.c.)\nonumber \\
-\mu n_d + \sum_{q}\omega_q\vec{h}_{q}^{\dagger} \cdot \vec{h}_{q} + I
\sum_{q}\vec{S}_{d}\cdot(\vec{h}_q + \vec{h}_{-q}^{\dagger}).
\label{3old}
\end{eqnarray}
A formal derivation of $H_{\text{EDMFT}}$ in the limit $d\rightarrow \infty$
is given in Appendix \ref{EDMFT_derivation}.  Here $\tilde d_\sigma^+$
is a projected fermion creation operator for the impurity orbital (the
original operator $\tilde c_{o\sigma}^+$ at the chosen
``impurity'' site $0$), $n_d = \sum_\sigma \tilde d_\sigma^+ \tilde
d_\sigma$ and $\vec S_d = \frac{1}{2} \sum_{\sigma,\sigma{'}} \tilde
d_\sigma^+ \vec\tau_{\sigma\sigma{'}} \tilde d_{\sigma{'}}$.  The
fermionic bath is represented by free fermion operators
$c_{k\sigma}^{\dagger}$, the bosonic bath by free boson operators
$h_{q\alpha}^{\dagger}$, $\alpha = 1,2,3$ with $\vec{h}_{q} = (h_{q1},
h_{q2}, h_{q3})$, and $\sum_{q} (\vec{h}_{q} + \vec h_{-q}^{\dagger})$
playing the role of a fluctuating local magnetic field.  The
excitation spectrum of the bath degrees of freedom, $E_k$ and
$\omega_q$, as well as the coupling constants $V$ and $I$ have to be
determined self-consistently by equating both the single-particle
Green's function and the spin susceptibility of the impurity model
$G_{imp}, \chi_{imp}$ with the local Green's function $G_{loc}$ and
the local susceptibility $\chi_{loc}$ of the lattice model,
\begin{eqnarray}
G_{imp,\sigma}(i\omega) \! &=& \!- \int_0^\beta d\tau e^{i\omega\tau}
\langle T_\tau \tilde d_\sigma (\tau)\tilde d_\sigma^+(0)\rangle
\stackrel{!}{=}G_{loc}(i\omega) \nonumber \\
\chi_{imp,\alpha}(i\omega)\! &=&\! \int_0^\beta d\tau
e^{i\omega\tau}\langle T_\tau S_\alpha (\tau) S_\alpha (0)\rangle
\stackrel{!}{=}\chi_{loc}(i\omega).
\label{4old}
\end{eqnarray}
The local $G$ and $\chi$ are obtained from their lattice counterparts
(\ref{2}), (\ref{3}), taking into account (\ref{4}), (\ref{5}), and by
summation over all momenta
\begin{equation}
G_{loc}(\omega) = \sum_{\vec k}G_{\vec k}(i\omega)\;\;,
\label{7}
\end{equation}
\begin{equation}
\chi_{loc}(\omega) = \sum_{\vec q}\chi_{\vec q}(i\omega).
\label{8}
\end{equation}

As shown in Appendix \ref{EDMFT_derivation}, the self-energies
$\Sigma$ and $M$ also characterize the impurity Green's functions:
\begin{equation}
G_{imp}(i\omega) = \left[i\omega + \mu - V^2G_{c}(i\omega)
- \Sigma(i\omega)\right]^{-1}\;\;\;,
\label{9}
\end{equation}
\begin{equation}
\chi_{imp}(i\omega) = \left[M - I^2G_h\right]^{-1}\;\;\;,
\label{10}
\end{equation}
where
\begin{equation}
G_{c}(i\omega) = \sum_{k}\frac{1}{i\omega - E_{k}}\;\;,
\label{11}
\end{equation}
\begin{equation}
G_{h}(i\omega) = \sum_{q}
\frac{2\omega_{q}}{(i\omega)^2 - (\omega_{q})^2}\;\;\;,
\label{12}
\end{equation}
so that the system of equations (\ref{2}) - (\ref{10}) is closed.
It follows from (\ref{9},{10}) that only the densities of states
of the baths,
\begin{equation}
A_{c}(\omega) = \frac{V^2}{\pi}{\rm Im} \;\;G_{c}(\omega - i0)
= V^2\sum_{k}\delta (\omega - E_{k})
\label{13}
\end{equation}
and
\begin{eqnarray}
D_{h}(\omega) &=& \frac{I^2}{\pi} {\rm Im}\;\; G_{h}(\omega
- i0)\nonumber\\
&=& I^2\sum_{q} \Big[\delta (\omega - \omega_{q}) - \delta
(\omega + \omega_{q})\Big]\;\;,
\label{14}
\end{eqnarray}
are needed.  For practical purposes we have included the coupling
constants $V$ and $I$, respectively, in the definitions of the density
of states.

\section{Generalized Non-Crossing Approximation}
\label{s.NCA}

The solution of the quantum impurity model (\ref{3old}) for given
$A_c(\omega)$ and $D_h(\omega)$ is difficult.  Many of the methods
developed in the past for solving Anderson impurity models in the
context of DMFT such as iterated perturbation theory \cite{GGKR} and
the numerical renormalization group method \cite{Bu} are not applicable
in the case of a bosonic bath.  The Quantum Monte Carlo method has
been successfully applied to an anisotropic Kondo lattice model with
Ising-type spin coupling \cite{GS,PK}, but it is extremely difficult
to treat Heisenberg couplings with manageable effort. The only method
left to us is self-consistent perturbation theory like the
non-crossing approximation (NCA) or the conserving T-matrix
approximation (CTMA) \cite{KW,B}.

We will therefore employ a conserving diagrammatic approximation in
which infinite classes of perturbation theory in $V$ and $I$ are
resummed.  We are aiming at a level of approximation corresponding to
NCA for the usual Anderson model.  A convenient way to phrase the
perturbation theory in the hopping $V$ and the exchange coupling $I$,
in the presence of an infinitely strong Coulomb repulsion $U$, is in
terms of a pseudo-particle representation.  We define pseudo-fermion
operators $f_\sigma^+, \sigma = \uparrow, \downarrow$, creating the
singly occupied impurity state and the slave boson operator $b^+$
creating the empty impurity level, when acting on a corresponding
vacuum state \cite{Barnes}.  Since the local level is either empty or
singly occupied, the operator constraint $Q = b^+ b + \sum_\sigma
f_\sigma^+f_\sigma = 1$ has to be satisfied at all times.  The
constraint is enforced exactly by adding a term $\lambda Q$ to the
Hamiltonian and taking the limit $\lambda \rightarrow \infty$ (see
(\ref{22}) below).  The projected local electron operators $\tilde
d_\sigma$ may then be replaced by $b^+f_\sigma$, turning the problem
into a many-body system of pseudo-fermions $f_\sigma$ and slave bosons
$b$, interacting with the fermions $c_{k\sigma}$ and bosons $\vec h_q$
of the bath.

It is essential for any approximation scheme to respect the projection
and  not to allow transitions between different sectors of Hilbert
space labeled by $Q$.  To this end we employ a conserving
approximation specified by a generating Luttinger-Ward type functional
$\Phi$ from which all self-energies are obtained as functional
derivatives, $\Sigma_a = \frac{\delta\Phi}{\delta G_a}$.  The building
blocks of $\Phi$ are the dressed Green's functions of pseudofermions
$G_f$ (depicted as dashed line), slave bosons $G_b$ (wiggly line),
bath fermions $G_c$ (solid line) and bath bosons $G_h$ (curly line)
and the vertices corresponding to hopping $V$, and exchange
interaction, $I$.

\begin{figure}
\begin{center}
\includegraphics[height=0.9\linewidth,angle=-90]{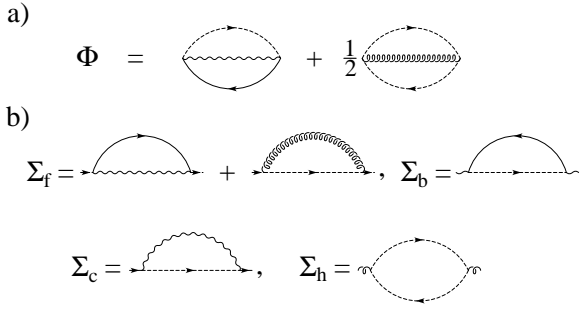}
\end{center}
\caption{\label{luttWard} The two lowest order contributions to the
Luttinger-Ward functional $\Phi$ and corresponding self-energies. Only
diagrams with no line-crossings are taken into account (a
generalization of NCA). The broken (wavy) line denotes pseudo fermion
(pseudo-boson) Green's function $G_{f}$ ($G_b$), and the solid
lines represent the conduction electron Green's functions
$G_{c}$, the curly line the correlator $G_{h}$ of the
bosonic bath. Also shown are the pseudo self-energies as well as
self-energies of the baths.}
\end{figure}

In the strongly incoherent regime we are interested in, vertex
corrections are not expected to change the behavior in a qualitative
way.  They may, however, lead to quantitative changes.  In this paper
we would like to explore the leading behavior first, so that we may
neglect vertex corrections for the moment.  The lowest order terms of
$\Phi$ in self-consistent perturbation theory in the bare coupling
constants $V$ and $I$ are shown in Fig.~\ref{luttWard}a. 
The first one is the known generating
functional of NCA, whereas the second one is new and involves the bath
bosons.  The corresponding self-energies are shown in
Fig.~\ref{luttWard}b, for the pseudofermions $(\Sigma_f)$ and slave
bosons $(\Sigma_b)$, as well as the bath fermions $(\Sigma_c)$ and the
bath bosons $(\Sigma_h)$.  We note that the impurity single-particle
Green's function after projection $(\lambda \rightarrow \infty)$ is
related to $\Sigma_c$ by \cite{CKW}
\begin{equation}
G_{imp}(i\omega)  = \frac{1}{V^2} \Sigma_c(i\omega)
\label{15}
\end{equation}
and likewise the impurity spin susceptibility is proportional to the
bath boson self-energy
\begin{equation}
\chi_{imp}(i\omega) = -\frac{1}{I^2}\Sigma_h(i\omega).
\label{16}
\end{equation}
Cutting a pseudofermion line in each of the two diagrams of the
generating functional, Fig.~\ref{luttWard}a, one finds two diagrams
for the pseudofermion self-energy
\begin{equation}
\Sigma_{f}(i\omega) = \Sigma_{f}^{(2a)}(i\omega) + \Sigma_{f}^{(2b)}(i\omega)
\label{17}
\end{equation}
as depicted in Fig.~\ref{luttWard}b.  Likewise, the slave boson
self-energy $\Sigma_b$, the fermion bath self-energy $\Sigma_c$ and
the boson bath self-energy $\Sigma_h$ are obtained by cutting the
respective Green's function lines in the two diagrams of $\Phi$.  The
corresponding analytical expressions are given by
\begin{subequations}
\label{18abcde}
\begin{align}
\Sigma_{f\sigma}^{(2a)}(i\omega) &= - V^2 T
\sum_{\omega{'}}G_{c\sigma}(i\omega{'}) G_b(i\omega - i
\omega{'})\;\;,\\ 
\Sigma_{f\sigma}^{(2b)}(i\omega) &= - \frac{1}{4}I^2
\sum_{\sigma{'},\alpha}\tau_{\sigma\sigma{'}}^\alpha
\tau_{\sigma{'}\sigma}^\alpha  \nonumber\\
& \qquad\quad \times T \sum_\Omega G_{h\alpha}(i\Omega)G_{f\sigma{'}}(i\omega + i\Omega)\;\;,\\ 
\Sigma_b(i\Omega) &= V^2T
\sum_{\sigma,\omega{'}}G_{c\sigma}(i\omega{'})G_{f\sigma}(i\Omega + i
\omega{'})\;\;,\\ 
\Sigma_{c\sigma}(i\omega) &= - V^2T \sum_\Omega G_{f\sigma}(i\omega +
i\Omega)G_b(i\Omega)\;\;,\\ 
\Sigma_{h\alpha}(i\Omega) &= \frac{1}{4}I^2
\sum_{\sigma,\sigma{'}}\tau_{\sigma\sigma{'}}^\alpha
\tau_{\sigma{'}\sigma}^\alpha \nonumber\\
&\qquad\quad \times T\sum_{\omega{'}}G_{f\sigma}(i\omega{'})
G_{f\sigma{'}}(i\omega{'}+i\Omega)\;\;,
\end{align}
\end{subequations}
\noindent 
where $i\omega, i\omega{'}$ and $i\Omega$ are fermionic and bosonic
Matsubara frequencies, respectively; $\sigma,\sigma{'} = \uparrow,
\downarrow$; $\alpha = 1,2,3$.  Next one may transform the Matsubara
frequency sums into frequency integrals along the branch-cuts of the
Green's functions and perform the analytical continuation to the real
frequency axis.  The projection to the singly-occupied sector of
Hilbert space may now be carried out.  To this end the frequency
arguments of the pseudoparticle Green's functions are shifted by the
chemical potential $\lambda$ and the limit $\lambda \rightarrow
\infty$ is taken.  This yields
\begin{subequations}
\label{19abc}
\begin{align}
\Sigma_{f\sigma}^{(2a)} (\omega + i0) &= \int d\xi
f(-\xi)A_{c\sigma}(\xi)G_b(\omega - \xi + i0)\;\;,\\ 
\Sigma_{f\sigma}^{(2b)}(\omega + i0) &= \frac{1}{4}
\sum_{\sigma{'},\alpha}\tau_{\sigma\sigma{'}}^\alpha
\tau_{\sigma{'}\sigma}^\alpha  \nonumber\\
&\times\int d\xi n(\xi)D_{h\alpha}(\xi)G_{f\sigma{'}}
(\omega+\xi + i0)\;\;,\\
\Sigma_b(\omega + i0) &= \sum_\sigma \int d\xi
f(\xi)A_{c\sigma}(\xi)G_{f\sigma}(\omega + \xi + i0).
\end{align}
\end{subequations}
where $f(\xi)$ and $n(\xi)$ are the Fermi and Bose functions,
respectively, and $A_c(\xi)$ and $D_h(\xi)$ are spectral functions of
the fermionic and bosonic baths as defined in (\ref{13}), (\ref{14}).

Since we incorporated the factors of $V^2$ and $I^2$ into the
definition, $A_{c}$ and $D_h$ are not normalized anymore, their
total weight being given by $V^2$ and $I^2$, respectively.

The projected pseudoparticle Green's functions are expressed in terms
of their self-energies as
\begin{subequations}
\label{20ab}
\begin{align}
G_{f}(\omega + i0) & = \frac{1}{\omega + \mu - \lambda_0 -
\Sigma_{f}(\omega + i0)}\;\;,\\
G_b(\omega + i0) & = \frac{1}{\omega - \lambda_0 - \Sigma_b(\omega+ i0)}\;\;,
\end{align}
\end{subequations}
where the (finite) energy shift $\lambda_0$ is determined by fixing the 
local charge $Q$, \cite{CKW}
\begin{eqnarray}
&\lim_{\lambda\to\infty} e^{\beta\lambda}\left\langle \sum_\sigma
f_\sigma^+f_\sigma + b^+b \right\rangle_G\nonumber\\
&= \int d\omega e^{-\beta\omega}
\Big[\sum_\sigma A_{f\sigma}(\omega) + A_b(\omega)\Big]=1 .
\label{21}
\end{eqnarray}
Here the subscript $G$ specifies an expectation value in the grand
canonical ensemble and $A_{f}(\omega) = - \frac{1}{\pi} {\rm Im}
G_{f}(\omega + i0)$, etc.

The remaining self-energies $\Sigma_c$ and $\Sigma_h$ contain one
pseudoparticle loop each and are therefore $\propto
e^{-\beta\lambda}$.  The projected expectation value of any operator
that vanishes in the $Q = 0$ subspace is then given by \cite{AA}
\begin{equation}
\langle A\rangle = \lim_{\lambda\to\infty} \frac{\langle A
\rangle_G}{\langle Q\rangle_G} = \lim_{\lambda\to\infty}
e^{\beta\lambda} \langle A \rangle_G
\label{22}
\end{equation}
using (\ref{21}).  It follows that
\begin{widetext}
\begin{eqnarray}
\Sigma_{c,\sigma} (\omega + i0) = & V^2 \int d\xi
e^{-\beta\xi}\Big[G_{f\sigma}
(\xi + \omega + i0)A_b (\xi)  - A_{f\sigma}(\xi)G_b(\xi - \omega -
i0)\Big] .
\label{23}
\end{eqnarray}
With the help of (\ref{15}) we find the imaginary part of the impurity
Green's function in the compact form
\begin{equation}
{\rm Im}\;\; G_{imp,\sigma} (\omega + i0) = - \frac{\pi}{f(-\omega)} \int
d\xi e^{-\beta\xi}A_{f\sigma}(\xi + \omega)A_b(\xi).
\label{24}
\end{equation}

From (\ref{18abcde}e) one finds after analytical continuation and projection
\begin{equation}
\Sigma_{h\alpha} (\omega + i0) = \frac{I^2}{4}\sum_{\sigma,\sigma{'}}
\tau_{\sigma\sigma{'}}^\alpha \tau_{\sigma{'}\sigma}^\alpha \int d\xi
e^{-\beta\xi} \Big[A_{f\sigma}(\xi)G_{f\sigma{'}}(\xi + \omega + i0) +
G_{f\sigma}(\xi - \omega - i0)A_{f\sigma{'}}(\xi)\Big].
\label{25}
\end{equation}
The impurity susceptibility is obtained from (\ref{16},\ref{25}) as
\begin{equation}
{\rm Im}\;\;\chi_{imp,\alpha}(\omega + i0) =
\frac{\pi}{4n(\omega)}\sum_{\sigma,\sigma{'}}\tau_{\sigma\sigma{'}}^\alpha
\tau_{\sigma{'}\sigma}^\alpha\int d\xi e^{-\beta\xi}A_{f\sigma}(\xi -
\omega) A_{f\sigma{'}}(\xi).
\label{26}
\end{equation}
\end{widetext}
Equations (\ref{4old})-(\ref{12}), together with the ``impurity solver'',
Eqs.~(\ref{17},{19abc})-(\ref{21},\ref{24},\ref{26}) have been
solved self-consistently.  Starting with given initial values of the
fermionic and bosonic bath and pseudoparticle spectral functions,
$A_c(\xi)$ and $D_h(\xi)$, the first approximation to the pair of
impurity Green's functions $G_{loc}$ and $\chi_{loc}$ as
well as the pseudoparticle spectral functions is determined.
Using the identities
\begin{eqnarray}
G_{loc} = \sum_{\vec k}{1\over G_{loc}^{-1}+V^2 G_c
-\epsilon_{\vec k}}\\
\chi_{loc} = \sum_{\vec q}{1\over \chi_{loc}^{-1}-I^2 G_h+J_{\vec
q}}
\end{eqnarray}
that follow from equations (\ref{2},\ref{4old},\ref{7},\ref{9})
and (\ref{3},\ref{4old},\ref{8},\ref{10}), the new bath
spectral funtions $A_c=-{1\over\pi}{\rm Im}V^2 G_c$ and
$D_h=-{1\over\pi}{\rm Im}I^2 G_h$ may be deduced.
With these and the updated pseudoparticle Green's functions one
determines new $G_{loc}, \chi_{loc}, G_f, G_b$ with the help of
the impurity solver. The iteration is continued, until convergence is
found to the desired level.  This process is found to converge well in
the temperature regime $T\ge 0.04 t$ using a nearest-neighbor
tight-binding model, where $t$ is the hopping amplitude.  At lower $T$
a solution could not be found any more.  In the following we will
present the results of the numerical evaluation, before  discussing in
detail the reasons for the breakdown of the solution in the low
temperature domain.

\section{Results}\label{s.results}

\subsection{Local spectral function: pseudogap and non-Fermi liquid physics}

\begin{figure}
\begin{center}
\includegraphics[width=0.9 \linewidth]{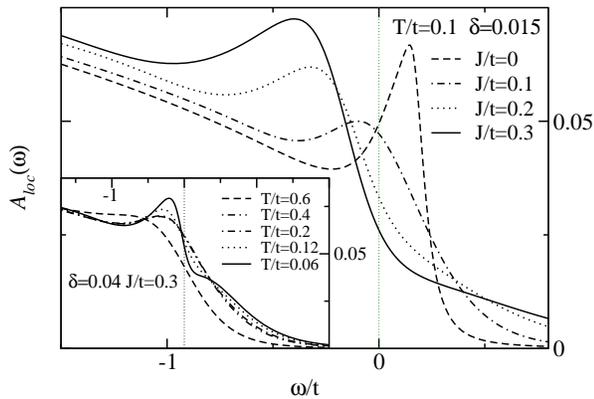}
\end{center}
\caption{\label{A00_J} The local spectral function plotted versus
frequency for four different $J/t=$ 0, 0.1, 0.2 and 0.3 and $T=0.1t$
for doping level of $\delta=0.015$. The evolution of the pseudogap of
width $J$ is clearly visible. The zero of energy is set at the
chemical potential $\mu$. The inset shows temperature dependence of
the local spectral function at the doping level 4\% and for $J=0.3t$.}
\end{figure}

\begin{figure}
\begin{center}
\includegraphics[width=0.9 \linewidth]{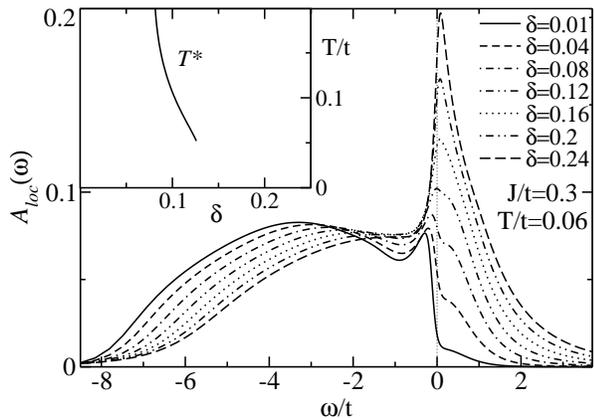}
\end{center}
\caption{\label{A00_T} The local spectral function plotted versus
frequency for $T=0.06t$ and $J/t=0.3$ for various hole-doping
concentrations $\delta$. The inset shows the characteristic
temperature $T^*$ where the pseudogap opens (for the definition see
the main text).}
\end{figure}

The most striking result of our work is the appearance of a pseudogap
in the local electron spectral function $A_{loc}(\omega)$ at small
hole doping and low temperatures.  Fig.~\ref{A00_J} shows how the
pseudogap starts to form when the exchange interaction $J$ is switched
on, for $\delta = 0.015$ and $T = 0.1t$.  In the limiting case of
$J=0$, corresponding to the Hubbard model in the limit $U \rightarrow
\infty$, $A_{loc}(\omega)$ is characterized by a broad maximum below
the Fermi level $(\omega = 0)$ interpreted as the lower Hubbard band,
and a narrow peak (``quasiparticle peak'') above $\omega = 0$.
  As $J$ is switched on, the quasiparticle peak
disappears rapidly and the weight under it appears to be shifted a
distance $\sim J$ below the Fermi level, forming a peak-dip-hump
structure.  The width of the pseudogap appears to scale with $J$.  At
the same time the spectral function develops a tail above $\omega = 0$
reaching far $(\sim t)$ above the bare band edge.  It is instructive
to observe how the pseudogap disappears for a given $J = 0.3t$ at $T =
0.06t$ with increasing doping level (Fig.~\ref{A00_T}).  The pseudogap
vanishes and the quasiparticle peak begins to appear at dopings above
$\delta \approx 0.1$.  We note in passing that the bulk of the
spectral weight in the lower Hubbard band is shifted rigidly with the
chemical potential and only a section of width $\sim 4 {\rm max} (J,
\delta t)$ at the chemical potential is changing with the doping.

The formation of the pseudogap at a low doping $\delta = 0.04$ and
fixed $J = 0.3t$ as the temperature is lowered from $T = 2J$ down to
$T = 0.2J$ is shown in the inset of Fig.~\ref{A00_J}.  In order to
quantify the appearance of the pseudogap for given $\delta$ as a
function of $T$ one may define the temperature $T^*$ at which the
curvature of $A_{loc}(\omega)$ at $\omega = 0$ changes sign from
negative to positive values as $T$ is lowered.  In the inset of
Fig.~\ref{A00_T} the $T^*$ values determined in this way are plotted
versus $\delta$.  $T^*$ is seen to drop rapidly with $\delta$, tending
to zero at $\delta \sim 0.15$. These results are reminiscent of what
is seen in ARPES experiments on high $T_c$ superconductors \cite{I}. 

\begin{figure}
\begin{center}
\includegraphics[width=0.9 \linewidth]{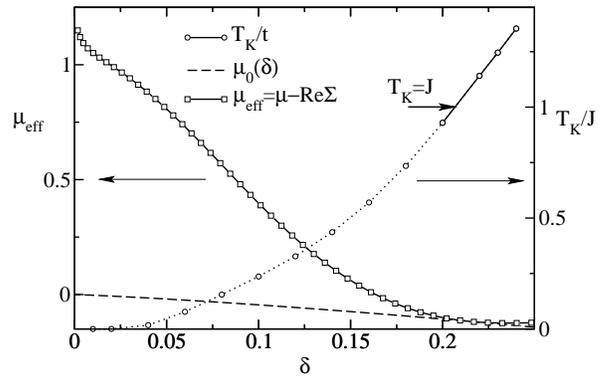}
\end{center}
\caption{\label{Tk} The effective chemical potential $\mu_{\text{eff}}$ vs
doping for $J=0.3t$ and $T=0.06t$ (left scale). The dotted line shows
the non-interacting chemical potential $\mu_0$.  Open circles mark the
estimation for the Kondo temperature $T_K$ vs doping as calculated
from Eq.~\ref{28} (right scale). The arrow marks the position where
$T_K$ is equal to $J$. Only in the regime where $T_K$ is larger than
$J$, the solution shows the onset of a Fermi liquid phase.}
\end{figure}

How is the pseudogap generated? The clue to this question lies in the
behavior of the effective chemical potential $\mu_{\text{eff}} = \mu - {\rm
  Re}\Sigma (0)$, as a function of doping.  In Fig.~\ref{Tk}, $\mu_{\text{eff}}$
is shown at a low temperature $T = 0.06 t$, in comparison with the
bare chemical potential $\mu_0$ (of the noninteracting system).  At
doping levels $\delta \gtrsim 0.2$ one finds that $\mu_{\text{eff}}$
coincides namely
with $\mu_0$, a necessary condition for Fermi liquid behavior.  Upon
lowering the doping concentration, $\mu_{\text{eff}}$ is seen to grow until
at $\delta \approx 0.02$ the upper edge of the bare band is reached
(the zero of energy is fixed at the center of the tight-binding band).
In fact $\mu_{\text{eff}}$ moves above the bare band, signalling the
availability of states even above the latter.  By contrast, the bare
chemical potential remains located in the center of the band,
approaching $\mu_0 = 0$ in the limit $\delta\rightarrow 0$.  The fact
that $\mu_{\text{eff}}$ is moving up towards the upper band edge for
$\delta\rightarrow 0$ is a strong and unequivocal signal of non-Fermi
liquid behavior -- it is only possible for a highly incoherent metal
with a large $\text{Im} \Sigma$.  It is interesting to recall that in
DMFT for the Hubbard model (which in the limit $U \rightarrow \infty$
is identical to the $t$-$J$ model for $J\rightarrow 0$) one finds Fermi
liquid behavior at low temperatures, and $\mu_{\text{eff}} = \mu_o$.  Even at
not so low temperatures $(T \gtrsim 0.06t)$ for $J=0$ $\mu_{\text{eff}}$
follows $\mu_0$ except at rather low doping values $\delta \lesssim
0.05$, where a strong temperature dependence appears.

Similar behavior has been found in Refs. [\onlinecite{Maier,Phillips}]
for the Hubbard model at intermediate coupling.  In Ref.
[\onlinecite{Maier}] a the dynamical cluster approximation involving
up to 64 sites was employed and the mean-field equations were solved
by QMC simulation and the maximum entropy method, to effect the
analytical continuation from imaginary to real frequencies.
Maier {\it et al.}\cite{Maier}
interpreted the pseudogap found in their spectra as generated by
finite range antiferromagnetic correlations on the cluster or as RVB
physics.  Since the results are so similar to ours, and within our
approach finite range AF correlations or the formation of intersite
singlets are not included, we suggest that their pseudogap is created
by the same mechanism we identify as being responsible for our
pseudogap: incoherent fluctuations (see above). Stanescu and
Phillips\cite{Phillips} used a two-site cluster approach to  derive
nonlocal DMFT equations.  The quantum impurity model was solved by an
adaptation of the non-crossing approximation.  Again the results for
the spectral functions are similar to ours.  The authors claim that an
effective low-energy model cannot be defined, as low and high-energy
sectors are mixed in a dynamical way.  We do not see any reason for
such an unusual situation, neither from their paper nor from outside
arguments.  Rather, in the limit $U\gg t$, or more precisely, if $U$
is strong enough to generate a Mott gap, the separation of the lower and
upper Hubbard band is well defined, and a projection onto the lower
band is justified.  In Ref.  [\onlinecite{Phillips}] the appearance of
the pseudogap is attributed to short-range (nearest-neighbor)
correlations, limiting the phase space for low-energy excitations.
These correlations are identified as orbital ring currents flowing
between three adjacent sites. Since such effects are not included in
our calculation, and we nonetheless find a pseudogap and a violation
of Luttinger's theorem, very similar to Ref.  [\onlinecite{Phillips}], we
conclude that the (and sketched above) interpretation given in Ref.
[\onlinecite{Phillips}], is not correct.

We conclude that the behavior found in our scheme for low doping, namely
pseudogap and non-Fermi liquid physics, is a generic feature of an
incoherent metal. We have found this incoherent state to be quite
robust, e.g. against changes in band structure.  It is worth
mentioning that Parcollet and Georges \cite{PG} recently studied a $t$-$J$
model with random $J$, which is equivalent to our EDMFT equations for
the Bethe lattice.  They did not find indications for a pseudogap.
We believe the reason is  that they employ slave boson mean-field
theory, and thereby miss the incoherent part of the spectral function.
A similar spin model has been considered before by Sachdev and Ye
\cite{SY}. 

\begin{figure}
\begin{center}
\includegraphics[width=0.9 \linewidth]{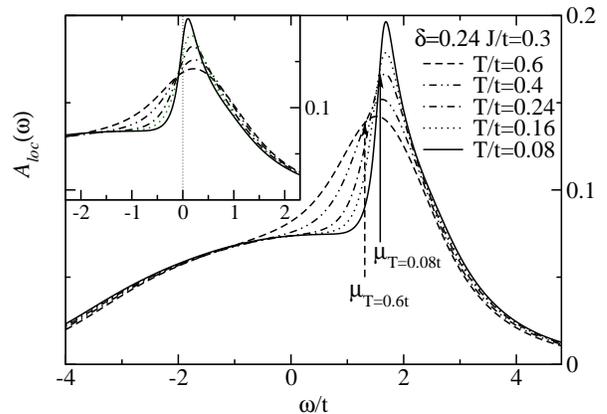}
\end{center}
\caption{\label{A24}
Temperature evolution of local spectral function for doping level
$\delta=0.24$. In the main panel, the arrows show the position of the
chemical potential while the inset shows spectra with the chemical
potential fixed at $\omega=0$. The evoultion of broad quasiparticle
peak above the Fermi level is clearly visible.}
\end{figure}

At larger dopings the solution shows the onset of a Fermi liquid
phase, which we now proceed to discuss.
First we show in Fig.~\ref{A24} the local spectral function at $\delta
= 0.24$ and $J = 0.3t$.  With increasing
temperature the quasiparticle peak broadens and the chemical potential
shifts to lower energies.  The value of $A(\omega)$ at the Fermi level
increases with falling temperature and tends to a limiting value as $T
\rightarrow 0$.

At  large doping the exchange interaction $J$ is 
unimportant and the EDMFT model reduces to an Anderson impurity model.
We may estimate the hybridization width $\Gamma$ of this model from
the density of states of the fermionic bath at the Fermi level
$(\omega = 0)$:
\begin{equation}
\Gamma = \pi A_c(\omega = 0).
\label{27}
\end{equation}
The energy of the local orbitals $E_d$, according to (\ref{3old}). is
given by the chemical potential $E_d = - \mu$.  An estimate of the
Kondo temperature is obtained from the conventional expression $T_K =
\sqrt{D\Gamma} \exp (\frac{\pi E_d}{2\Gamma})$ as
\begin{equation}
T_K = \sqrt{D\pi A_c(0)}\exp (-\frac{\mu}{2A_c(0)}).
\label{28}
\end{equation}
Fig.~\ref{Tk} shows $T_K/t$ as a function of $\delta$ for the low
temperature $T = 0.1t$, using $D = 2t$.  The Kondo temperature is
seen to fall strongly with decreasing $\delta$ even at the highest
value $\delta = 0.24$, and approaches zero rapidly in the pseudogap
regime.  The value where $T_K = J$ is indicated.  In the regime $T_K
\lesssim J$ one expects the exchange interaction to be of increasing
importance, such that the interpretation in terms of an Anderson
impurity model loses its meaning.  

\begin{figure}
\begin{center}
\includegraphics[width=0.9 \linewidth]{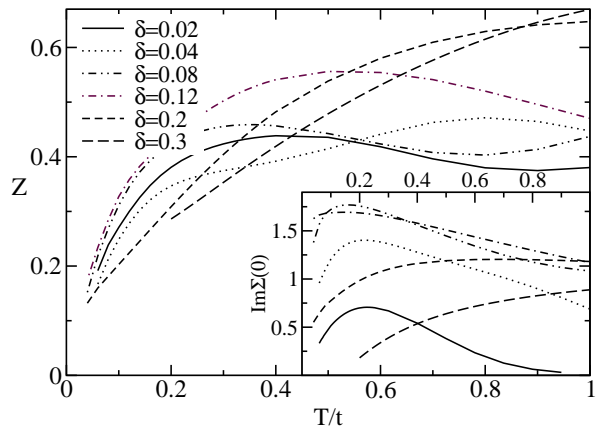}
\end{center}
\caption{\label{ImS0} Quasiparticle renormalization amplitude $Z$
plotted vs temperature for various doping concentrations. The inset
shows the imaginary part of the self-energy at zero frequency as a function
of temperature.}
\end{figure}

In the Fermi liquid regime the imaginary part of the self-energy of
$G_{loc}$ is expected to vary as
\begin{equation}
{\rm Im}\;\;\Sigma (\omega - i0) \sim t \Big[\omega^2 + (\pi T)^2\Big]/T_K^2
\label{31}
\end{equation}
where the Kondo temperature $T_K$ plays the role of the renormalized
Fermi energy.  The quadratic dependence is expected to hold for
$\omega, T \ll T_K$. The inset of Fig.~\ref{ImS0} shows ${\rm
Im}\;\Sigma (0)$ as a function of $T$ for doping levels from $\delta =
0.3$ down to $0.01$.  A limiting quadratic temperature dependence is not
seen since the lowest temperature reached in our evaluation is above $T_K$
(or, for $\delta > 0.2$, only slightly below $T_K$).

However, for $\delta = 0.24$
and $0.3$  behavior consistent with Fermi liquid theory would smoothly
match the results shown; For smaller doping, in particular around
$\delta \sim 0.1$, ${\rm Im} \;\Sigma$ at $T \sim 0.03 t$ is so large
that it is impossible to connect this behavior smoothly with a Fermi
liquid behavior below $T_K \simeq 0.1 t$.  At still smaller
$\delta\;\; {\rm Im}\;\Sigma$ is seen to decrease with doping, due to
the formation of the pseudogap.

As a further indication of Fermi liquid behavior we evaluate the quasiparticle
weight factor $Z$ defined as
\begin{equation}
Z = \Big( 1- \frac{\partial {\rm Re}\Sigma}{\partial
\omega}\Big)_{\omega = 0}^{-1}.
\label{32}
\end{equation}
Fig.~\ref{ImS0} shows $Z$ as a function of temperature for $\delta =
0.02 - 0.3$.  A finite quasiparticle weight in the limit $T
\rightarrow 0$ would signal Fermi liquid behavior.  It is seen that
only for the highest doping levels $\delta = 0.3$ and $0.24$ would an
extrapolation to $T = 0$ give a finite value.  For smaller values of
$\delta$ the $Z$-factor appears to drop rapidly with decreasing
temperature, possibly extrapolating to zero.

\subsection{Pseudoparticle spectral functions}

\begin{figure}
\begin{center}
\includegraphics[width=0.9 \linewidth]{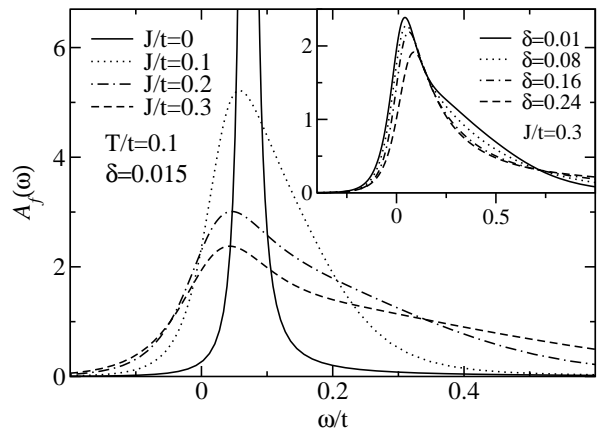}
\end{center}
\caption{\label{Spcf} The pseudo-fermion spectral function plotted vs
frequency for four different values of $J$. The inset shows the
evolution of spectra by doping the system at constant $J=0.3t$.  }
\end{figure}
The pseudofermion spectral function $A_f(\omega)$ at $\delta = 0.
015$ and $T = 0.1t$ is plotted versus $\omega/t$ in Fig.~\ref{Spcf},
for values of $J/t$ from $0$ to $0.3$.  While at $J = 0$ $A_f(\omega)$
is characterized by a narrow peak at a frequency $\omega \sim T$ and
of width $\sim T$, increasing $J$ leads to a rapid asymmetric
broadening of the peak, of width $\Delta \omega \sim J$.  Although in
the limit $T\rightarrow 0$ for general reasons one expects
$A_f(\omega)$ (and also $A_b(\omega)$) to acquire power-law divergent
behavior at the threshold $\omega = 0$  \cite{CSKW}, the temperature $T
= 0.1t$ is too high to show the asymptotic behavior.  At large doping,
$\delta > 0.2$, when the Kondo temperature $T_K$ as defined in
(\ref{28}) is larger than $J$, $A_f(\omega)$ is hardly affected by
$J$.  The doping dependence of $A_f(\omega)$ at $J = 0.3t$, as shown
in the inset of Fig.~\ref{Spcf}, is weak.  The characteristic energy
scale is $\max( J, T_K)\approx J$ up to the highest doping of
$\delta = 0.24$, and hence is independent of $\delta$.

\begin{figure}
\begin{center}
\includegraphics[width=0.9 \linewidth]{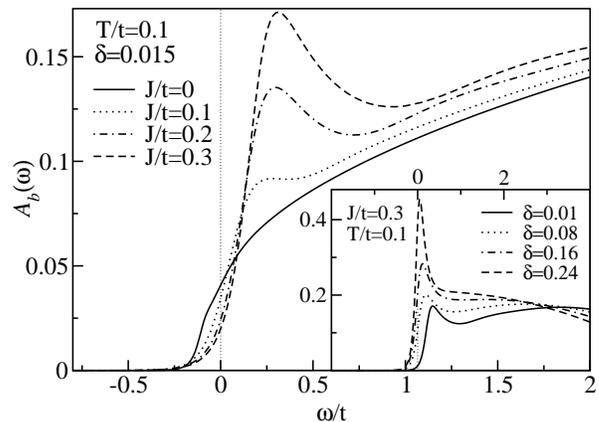}
\end{center}
\caption{\label{Spcb} The pseudo-boson spectral function for the same
parameters as used in Fig.~\ref{Spcf}.  }
\end{figure}
The pseudoboson spectral function shown in Fig.~\ref{Spcb} is roughly
speaking a mirror image of the lower Hubbard band.  As $J$ is
switched on spectral weight is pushed from below the threshold at
$\omega = 0$ and from the far end of the Hubbard band into a peak at
$\omega \sim J$, emulating the peak-dip-hump structure in
$A_{loc}(\omega)$ in the pseudogap regime.

Both in the case of $J=0$ and for $\delta > 0.2$ a sharp
quasiparticle peak is observed to form in $A_b(\omega)$ at $\omega =
0$.  The peak is suppressed at temperatures $T\gg T_K$, which is why
it is not seen in Fig.~\ref{Spcb}.  In contrast to $A_f(\omega)$,
$A_b(\omega)$ is strongly doping dependent, as shown in
Fig.~\ref{Spcb}.  At the moderately low temperature of $T = 0.1t$ and
for $J = 0.3t$, the pseudogap feature at small $\delta$ is seen to
smoothly cross over to the quasiparticle peak at large doping.

\subsection{Dynamical spin susceptibility}

\begin{figure}
\begin{center}
\includegraphics[width=0.9 \linewidth]{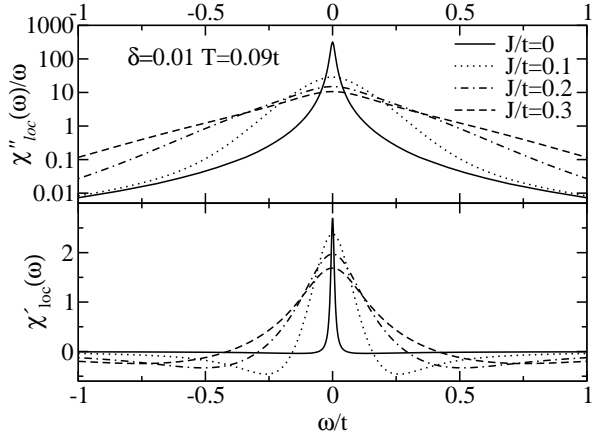}
\end{center}
\caption{\label{Chi_local}
The local dynamical spin susceptibility  plotted vs frequency for
four different $J/t$ and doping concentration $\delta=0.01$.
}
\end{figure}
The dynamical spin susceptibility is expected to reveal how the character
of spin fluctuations dependends on doping and exchange coupling
constant $J$.  
In Fig.~\ref{Chi_local} the imaginary part of
$\chi_{loc}(\omega)/\omega$ is shown at low doping, $\delta = 0.01$,
and low temperature, $T = 0.09t$, for various values of $J/t$ ranging
from $0$ to $0.3$.  As $J$ is increased, the peak of ${\rm
Im}(\omega)\chi/{\omega}$ broadens and the width is seen to be given by
$\Delta\omega \approx J$.  The real part ${\rm Re}\chi (\omega = 0) =
\chi{'}(0)$  decreases with $J$ as shown in Fig.~\ref{Chi_local}.
However, there is no trace of a pseudogap in ${\rm
Im}\chi_{loc}(\omega)$.
\begin{figure}
\begin{center}
\includegraphics[width=0.9 \linewidth]{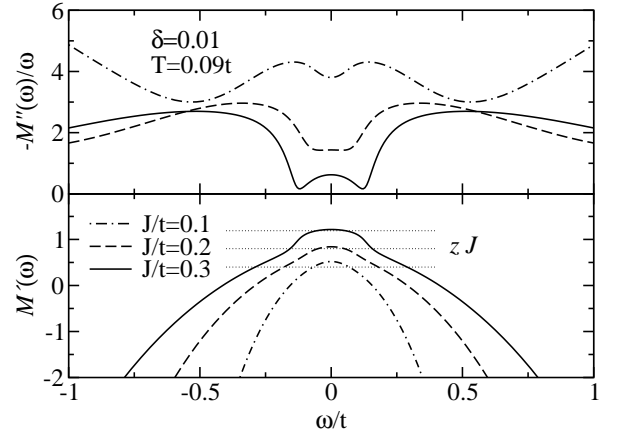}
\end{center}
\caption{\label{Mw}
The spin self-energy $M(\omega)$  plotted vs frequency for four
different $J/t$. The horizontal dotted lines mark the value $z J$, where
$z=4$ is the coordination number. 
}
\end{figure}
The pseudogap reveals itself in the spectrum of the self-energy 
$M(\omega)$ of magnetic excitations, as shown in Fig.~\ref{Mw}, where ${\rm Im}M(\omega) =
M''(\omega)$ is observed to develop a gap for $ \omega \lesssim J$. As
analyzed in Section \ref{instab.section}, the pseudogap is caused by large values of
$\chi{'}(\omega)$, which force a redistribution of spectral weight in
$M''(\omega)$ by way of the self-consistent feedback of
$\chi{'}(\omega)$ into $M''(\omega)$.
\begin{figure}
\begin{center}
\includegraphics[width=0.9 \linewidth]{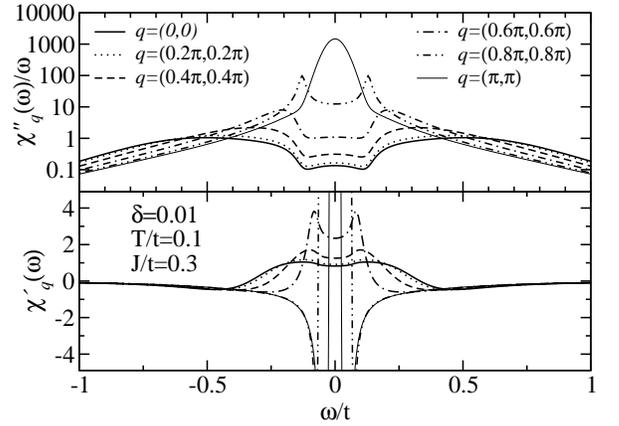}
\end{center}
\caption{\label{Chi_q}
The momentum dependent spin susceptibility along the $(0,0)- (\pi,\pi)$ axis  plotted as a function of
frequency for doping concentration $\delta=0.01$ and temperature
$T=0.1t$.}
\end{figure}
In Fig.~\ref{Chi_q}, the momentum-resolved spin excitation spectrum,
$\chi''_q(\omega)/\omega$ is shown for $J = 0.3t$, $\delta = 0.01$ and
$T = 0.1t$. Whereas a pronounced gap exists at $q$-values away from the
antiferromagnetic wavevector $Q = (\pi, \pi)$, near $Q$ the gap is
filled in.  This is due to the fact that in the region of $\vec
q$-space around $\vec Q$ not only $ M''(\omega)$ is small for
$\omega \lesssim J$, but also the real part of the denominator of
$\chi_q(\omega)$ vanishes, as $M{'}(\omega) + J_q
\rightarrow 0$ for $q \rightarrow Q$ and as the transition to the
antiferromagnetically ordered state is approached.  Consequently, the
ratio $M''(\omega)/| \chi_q(\omega)|^2$ develops
pronounced peaks at $| \omega | \sim J$ rather than a pseudogap.  
In the local susceptibility the contribution from $q \approx Q$
tends to fill in the pseudogap, which is therefore not discernible in
Fig.~\ref{Chi_local}. The effect of approaching the ordered state is
also observed in the real part of $\chi_q(\omega)$, shown in
Fig.~\ref{Chi_q}.  The static $q$-dependent susceptibility $\chi_q(0)$
is seen to grow by two orders of magnitude as $q$ is varied from $q =
0$ to $q = Q$.
\begin{figure}
\begin{center}
\includegraphics[width=0.9\linewidth]{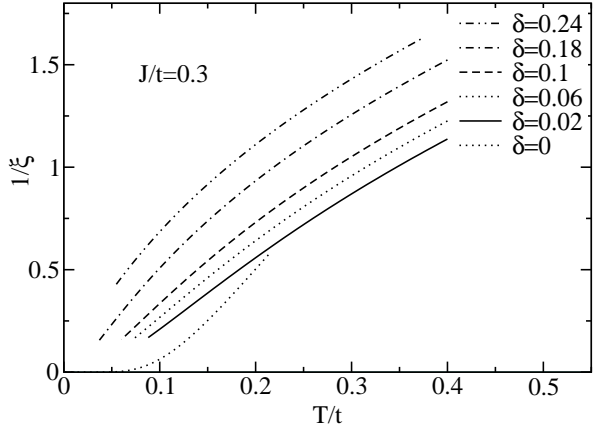}
\end{center}
\caption{\label{corrl}
The inverse of dynamic correlation length plotted as a function of
temperature for various doping levels. The
curve for $\delta=0$ is taken from Ref.~[\onlinecite{A1}] and
corresponds to the two-dimensional Heisenberg model.
}
\end{figure}
This behavior reflects the effect of a large spin correlation length
$\xi$, defined through
\begin{equation}
\chi_q(0) = \frac{2}{zJ}\frac{1}{\xi^{-2} + (\vec q - \vec Q)^2},
\label{33}
\end{equation}
for $\vec q \approx \vec Q$ ($z = 4$ is the coordination number and
length is measured in units of the lattice constant).  In
Fig.~\ref{corrl} the inverse correlation length is plotted versus
$T/t$ for $J = 0.3t$ and for various doping levels.  For comparison,
the theoretical  result  for the Heisenberg model (two loop order RG
of the nonlinear sigma model)
given in Ref. [\onlinecite{SHN,A1}] (limit $\delta = 0$)
is shown as well.  It appears to connect smoothly to the curve for
$\delta = 0.02$. Fig. \ref{corrl} also serves to show that the numerical
solution ceases to exist at $\chi_Q (0) \gtrsim 10^2$, as will be
discussed in Section \ref{instab.section}.

  \begin{figure}
  \begin{center}
  \includegraphics[width=0.9 \linewidth]{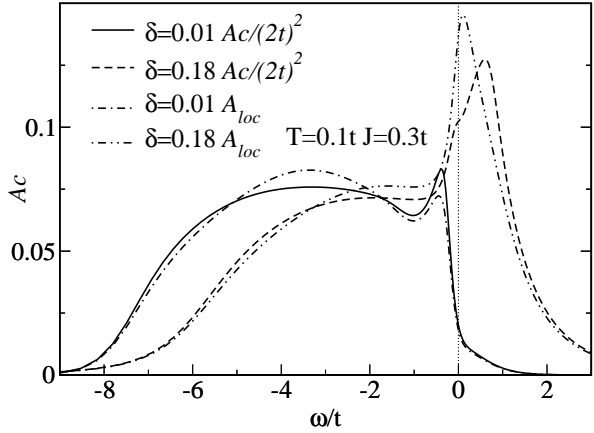}
  \end{center}
  \caption{\label{Ac}
  The fermionic bath spectral function $A_c$ for two different doping
  levels $\delta=0.01$ and $\delta=0.18$ at $J=0.3t$ and $T=0.1t$. The
  local spectral function is also shown for comparison.
  }
  \end{figure}

\subsection{Spectral functions of the fermionic and bosonic baths}
The spectral function $A_c(\omega)$ of the fermionic bath is shown for
$J = 0.3t$ and at $T = 0.1t$ in Fig.~\ref{Ac}.  The imposed
self-consistency of the EDMFT equations has led to a drastic
renormalization of the structureless tight-binding density of states.
In fact $A_c(\omega)$ reflects the structure seen in $A_{loc}(\omega)$
to a large degree: On the one hand, the quasiparticle peak at large
doping and on the other the pseudogap at small $\delta$.  For
comparison we show $A_{loc}(\omega)$ in Fig.~\ref{Ac} as well.

\begin{figure}
\begin{center}
\includegraphics[width=0.85\linewidth,clip=]{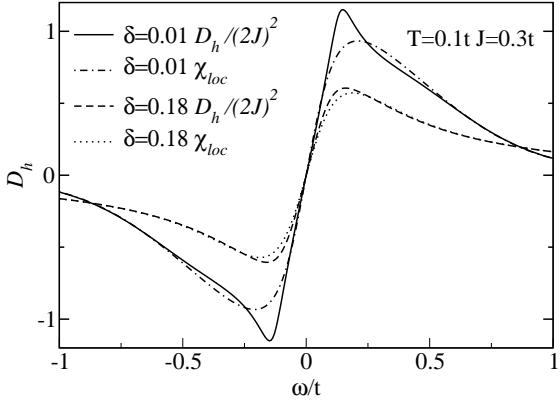}
\end{center}
\caption{\label{Dh} The bosonic bath spectral function $D_h$ for two
different doping levels $\delta=0.01$ and $\delta=0.18$ at $J=0.3t$
and $T=0.1t$. For comaparison, the local dynamic spin susceptibility
is also shown.}
\end{figure}

A similar trend is seen in the case of the spectral function of the
bosonic bath $D_h(\omega)$, as is apparent from Fig.~\ref{Dh}.  Here we
also see from the comparison with $\chi_{loc}''(\omega)$ a large
degree of similarity.

The total weight under the spectral functions $A_c(\omega)$ and
$D_h(\omega)$ is equal to the squares of the coupling constants $V^2$
and $I^2$, respectively.  As shown in Appendix \ref{sumrule}, the coupling
constant $V^2$ is fixed by sum rules and is given by
\begin{equation}
V^2 =  2t^2 (1+\delta).
\label{B1}
\end{equation}
In contrast, a similarly simple relation does not hold for $I^2$.
However,
$I^2$ may be related to $\chi_{loc}$ and $M$ as follows:
\begin{equation}
I^2 = \int_0^\infty \frac{d\omega}{\pi} {\rm Im}\!\left[M(\omega - i0) -
\chi_{loc}^{-1}(\omega - i0)\right].
\label{B2}
\end{equation}
It turns out that the numerical evaluation yields
\begin{equation}
I^2 \approx 2J^2 (1-\delta).
\label{B3}
\end{equation}
The first moment of the eigenfrequencies $\omega_q$ of the bosonic
bath, is given by the f-sum rule,
\begin{equation}
\bar{\omega}_q \equiv
\sum_q\omega_q =
\frac{\left\langle\epsilon^2\right\rangle}{2I^2}
\int_{-\infty}^{\infty} \frac{d\omega}{\pi} \omega \chi_{loc}''
(\omega),
\label{B4}
\end{equation}
where $\left\langle\epsilon^2\right\rangle = \int d\epsilon \epsilon^2 N_J(\epsilon)$ and
$N_J(\epsilon)$ is the density of states (DOS) of $J_{\vec q}$.

\subsection{Thermodynamic properties}

The thermodynamic potential $\Omega$ within EDMFT can be expressed in
terms of the impurity free energy $\Omega_{imp}$ and contributions
from the fermionic and bosonic baths:

\begin{widetext}
\begin{equation}
\Omega = \Omega_{imp} + k_B T \sum_{i\omega} \left\{\sum_{\vec
k,\sigma} \ln \left[G_{\vec
k\sigma}(i\omega)/G_{loc,\sigma}(i\omega)\right] + \frac{1}{2}\sum_{\vec q,\alpha} \ln \left[\chi_{\vec
q}^{\alpha\alpha}(i\omega)/\chi_{loc}^{\alpha\alpha}
(i\omega)\right]\right\}e^{i\omega 0^+}.
\label{34}
\end{equation}

Performing the analytical continuation from imaginary frequencies to
the real axis and expressing the momentum summations as energy
integrals, (\ref{34}) may be written as
\begin{equation}
\Omega = \Omega_{imp} + \frac{1}{\pi}\int d\epsilon
D(\epsilon){\rm Im}\left\{2\int d\omega f(\omega)\ln
\left[G_{loc}(\omega)(\omega + \mu -
\Sigma(\omega)-\epsilon)\right] + \frac{3}{2} \int d\omega\,  n(\omega)\ln
\left[\chi_{loc}(\omega)(M(\omega) + \frac{J}{t}\epsilon)\right]\right\}.
\label{35}
\end{equation}
\end{widetext}

The impurity free energy is given by the shift of the chemical
potential\cite{CKW}, $\lambda_0$, defined by (\ref{21})
\begin{equation}
\Omega_{imp} = \lambda_0.
\label{36}
\end{equation}
\begin{figure}
\begin{center}
\includegraphics[width=0.85 \linewidth]{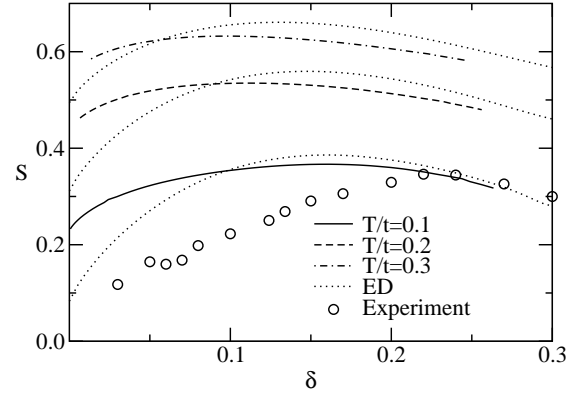}
\end{center}
\caption{\label{Entropy} Entropy per site as a function of doping
$\delta$ at various temperatures. Exact diagonalization results
\cite{JP} for the same temperatures are denoted by dotted lines while
the open circles correspond to the experimental data \cite{LMCAL} on
La$_{2-x}$Sr$_x$Cu0$_4$.}
\end{figure}
The entropy $S = -\Big(\frac{\partial\Omega}{\partial
T}\Big)_\mu$ as a function of doping concentration $\delta$ for
various temperatures is shown in Fig.~\ref{Entropy}.  Even at the low
temperature, $T = 0.1t$, $S$ is seen to be rather large $(\sim 0.5 \ln 2)$, an
indication for strong correlations and a rather incoherent state.  The
entropy of a noninteracting system at the same density would be about
an order of magnitude smaller.  The overall magnitude of $S$ compares
well with both the results of exact diagonalization \cite{JP} for a
small system and experimental data for La$_{2-x}$Sr$_x$Cu0$_4$ 
[\onlinecite{LMCAL}].  The calculated entropy shares the trend that it is
reduced both at large doping, when the system crosses over to a Fermi
liquid, and at smaller doping in the pseudogap phase.  The quenching
of the magnetic fluctuations by the incipient magnetic order as the
antiferromagnetic Mott insulator is approached for $\delta \rightarrow
0$ is qualitatively reproduced (note that for $J=0$, $S$ increases as
$\delta\rightarrow 0$, and this behavior is obtained in DMFT
calculations of the Hubbard model).

\begin{figure}
\begin{center}
\includegraphics[width=0.95 \linewidth]{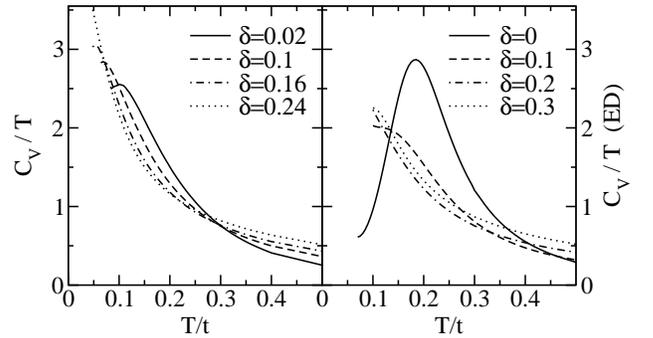}
\end{center}
\caption{\label{Cv} Specific heat coefficient vs
temperature for various doping concentrations. In the right panel we
show results obtained by the exact diagonalization \cite{JP}.}
\end{figure}
In Fig.~\ref{Cv} the specific heat divided by the temperature is
plotted versus $T$ for various dopings $\delta$ (left panel).  For
comparison the results of ED are shown on the right panel.

\begin{figure}
\begin{center}
\includegraphics[width=0.9 \linewidth]{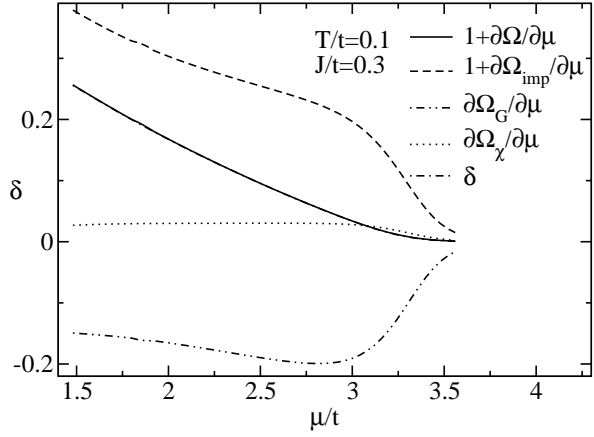}
\end{center}
\caption{\label{OmegaParts} Solid line: 
derivative of the thermodynamic
potential with respect to chemical potential
$1+(\partial\Omega/\partial\mu)_T$ (or equivalently doping vs. chemical potential).
The contributions from three different parts of the thermodynamic
potential: impurity, electron Green's function (second term in
Eq.~\ref{35}) and spin susceptibility part (last term in Eq.~\ref{35})
are shown separately.  }
\end{figure}
Another thermodynamic quantity of interest is the particle density
$n$, given by 
\begin{equation}
{n} = 1 - \delta = -\Big(\frac{\partial\Omega}{\partial\mu}\Big)_T.
\label{37}
\end{equation}
In Fig. \ref{OmegaParts} the doping $\delta$ is plotted versus $\mu$
at $T=0.1t$.  As expected, $\delta$ varies monotonically with $\mu$,
with positive curvature.

The particle density may also be obtained from the local Green's
function as $ {n} = 2 G_{loc,\sigma} (\tau = 0^+) $.  The resulting
values of $n$ are indistinguishable from those calculated by
differentiating $\Omega$,  which provides a check for numerical 
accuracy within our conserving approximation.

\subsection{Transport properties}

The calculation of transport properties in EDMFT is facilitated by the
observation that a momentum independent self-energy leads to a local
current vertex function (in other words, the non-local parts vanish in
the limit dimension $d\rightarrow \infty)$ \cite{IFT,M-H,K1}.  The
optical conductivity is therefore given by the single-particle Green's
function as
\begin{equation}
\sigma_{xx}(i\omega) = \frac{e^2}{\omega} k_B T \sum_{i\omega{'}}
\sum_{\vec k\sigma} (v_{\vec k}^x)^2 G_{\vec k}(i\omega{'})
G_{\vec k}(i\omega{'}+i\omega),
\label{40}
\end{equation}
where $v_{\vec k}^x = 2t \sin k_x$ is the bare current vertex. Using
the fact that $G_{\vec k}$ depends on $\vec k$ only through
$\epsilon_{\vec k}$ [see (\ref{2}), (\ref{4})], and performing the
analytical continuation to the real frequency axis one finds
\begin{eqnarray}
&&{\rm Re} \sigma_{xx} (\omega + i\delta) = 2\pi e^2 \int d\epsilon
\Phi_{xx}(\epsilon) \nonumber\\
&&\times \int d\omega{'} \frac{f(\omega{'})-f(\omega{'}+\omega)}{\omega}A(\epsilon,\omega{'})A(\epsilon,\omega{'}+\omega)\;\;,
\label{41}
\end{eqnarray}
where
\begin{equation}
\Phi_{xx}(\epsilon) = \sum_{\vec k}(v_{\vec k}^x)^2 \delta(\epsilon -
\epsilon_{\vec k})
\label{42}
\end{equation}
and $A(\epsilon_k,\omega) = \frac{1}{\pi}{\rm Im}G_{\vec k}(\omega -
i\delta)$.

Similarly, the off-diagonal or Hall conductivity in the presence of a
magnetic field ${\cal B}$ perpendicular to the plane takes the form
\cite{KY,VGJ,LK}
\begin{equation}
\sigma_{xy} = \frac{4\pi^2e^3}{3} {\cal B} \int d\epsilon \Phi_{xy}(\epsilon)\int
d\omega \Big(-\frac{\partial f}{\partial \omega}\Big)
\Big[A (\epsilon,\omega)\Big]^3\;\;,
\label{43}
\end{equation}
where 
\begin{equation}
\Phi_{xy} (\epsilon) = \sum_{\vec k} {\rm det}(\vec k) \delta
(\epsilon - \epsilon_{\vec k})
\label{44}
\end{equation}
and 
\begin{equation}
{\rm det}(\vec k) = \left| \begin{array}{llcl}
(\epsilon_k^x)^2 & \epsilon_k^{xy} \\
\epsilon_k^x\epsilon_k^y & \epsilon_k^{yy}\end{array}\right|\;\;;\;\;             
\epsilon_k^\alpha = \frac{\partial\epsilon_k}{\partial
k_\alpha}\;\;;\;\;\epsilon_k^{\alpha\beta} =
\frac{\partial^2\epsilon_k}{\partial k_\alpha\partial k_\beta}.
\label{45}
\end{equation}

\begin{figure}
\begin{center}
\includegraphics[width=0.9 \linewidth]{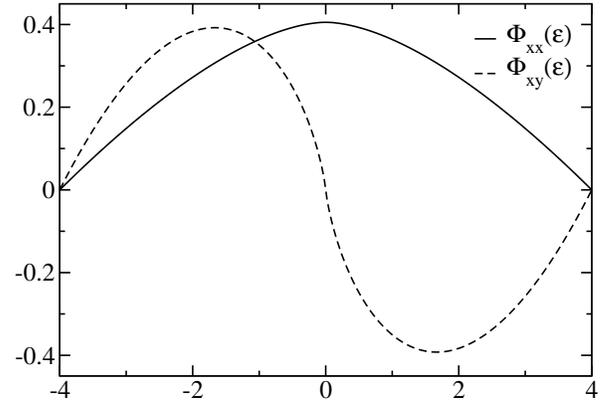}
\end{center}
\caption{\label{Doses}
The weigting functions for the two dimensional square lattice can be
expressed by elementary functions as
$\Phi_{xx}(x)={2t\over\pi^2}\left[2|x| E(1-1/x^2)\right.$ $\left.+2 K(1-1/x^2)-2\Pi(1-1/|x|,1-1/x^2)\right]$
and $\Phi_{xy}(x)= 2\left({2t\over\pi}\right)^2
[x^2 E(1-1/x^2)$ $-K(1-1/x^2))]{\rm Sign}(x)$.
Here, $K(x)$, $E(x)$ and $\Pi(x)$ are complete elliptic integrals of the
first, second and third kind and $x=\varepsilon/(4t)$.
}
\end{figure}
The weight factors $\Phi_{xx}$ and $\Phi_{xy}$ are shown in
Fig.~\ref{Doses}.  One observes that for the simple 2D tight-binding
lattice, $\Phi_{xx}$ is even function of energy while $\Phi_{xy}$ is
an odd function of energy.

The Hall coefficient ${\cal R}_H$ is defined as
\begin{equation}
{\cal R}_H = \frac{\sigma_{xy}}{\sigma_{xx}^2 {\cal B}}.
\label{46}
\end{equation}
For orientation it is useful to discuss the limit of low temperatures,
assuming ${\rm Im} \Sigma (\omega) \rightarrow 0$ at $\omega \rightarrow
0$, and $A(\epsilon,\omega)$  sharply peaked as a function of
$\epsilon$ at $\epsilon = \omega + \mu_{\text{eff}}$.  One may then do the
integrations on $\epsilon$ and $\omega$ in (\ref{41}) and (\ref{43}), yielding
\begin{equation}
\sigma_{xx} \simeq e^2 \frac{\Phi_{xx}(\mu_{\text{eff}})}{| {\rm Im}\Sigma (0)|}
\label{47}
\end{equation}
and
\begin{equation}
{\cal R}_H \simeq \frac{1}{2e} \frac{\Phi_{xy}(\mu_{\text{eff}})}
{[\Phi_{xx}(\mu_{\text{eff}})]^2},
\label{48}
\end{equation}
with $e=-|e|$.
We observe that ${\cal R}_H$ does not depend on ${\rm Im}\Sigma$ in this limit. In
the Fermi liquid regime $\mu_{\text{eff}} = \mu_0 < 0$, and consequently
$\Phi_{xy}(\mu_0) > 0$, leading to a negative $R_H < 0$.  

By contrast, in the incoherent regime of the $t$-$J$ model $\mu_{\text{eff}}$ is found to be positive, approaching the
upper band edge for $\delta \rightarrow 0$ (see Fig.~\ref{Tk}).  Since
$\Phi_{xy}(\epsilon)$ is negative for positive $\epsilon$, and
 ${\cal R}_H$ is seen to be positive
(hole-like).  For $\delta>0.17$, $\mu_{\text{eff}}$ changes sign and ${\cal R}_H$
turns negative. For the nearest neighbor tight-binding model, and
assuming a linear variation of $\mu_{\text{eff}}$ with $\delta$, $\mu_{\text{eff}} =
4t(1-C\delta)$, ${\cal R}_H$ takes the simple form
\begin{equation}
{\cal R}_H \simeq  \frac{\pi}{2C} \frac{1}{| e| \delta}\;\;;\;\;\delta
\rightarrow 0\;\;;\;\;C > 0.
\label{49}
\end{equation}
Using in addition the result for a single hole in the half-filled band
\cite{P_RH}, ${\cal R}_H = 1/| e|\delta$, one finds by comparison
$C = \frac{\pi}{2}$.  For the conductivity one obtains in a similar
way
\begin{equation}
\sigma_{xx} \simeq e^2 \frac{t\delta}{| {\rm
Im}\Sigma(0)|}\;\;\;,\;\;\delta\rightarrow 0.
\label{50}
\end{equation}
\begin{figure}
\begin{center}
\includegraphics[width=0.9 \linewidth]{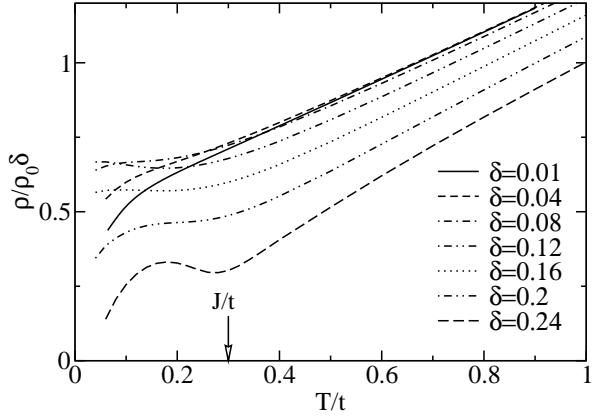}
\end{center}
\caption{\label{rho} $T$-dependence of the resistivity multiplied by
doping $\delta$. The linear $T$ behavior for high $T$ flattens for
$\delta>0.1$ at a temperature of the order of $J$. For $\delta<0.1$
the resistivity drops in the regime where a pseudogap opens.}
\end{figure}
Although (\ref{49}) and (\ref{50}) are in qualitative agreement with
our numerical results, we emphasize that the assumption of small ${\rm
  Im} \Sigma(0)$ is not justified in the incoherent regime.  A large
$\text{Im} \Sigma$ is actually necessary to obtain a $\mu_{\text{eff}}$ close
to the band edge and therefore a positive sign of $R_H$.

We now present the numerical results.  In Fig.~\ref{rho} the scaled
resistivity $\rho_{xx}\delta/\rho_0$, where $\rho_o = \hbar/e^2$, is
plotted versus temperature for values of $\delta$ ranging from $0.01$
to $0.23$.  The curves form a narrow band meaning that the scaling
$\rho_{xx}\propto 1/\delta$ shown in (\ref{50}) holds approximately
(and ${\rm Im}\Sigma(0)$ is a weak function of $\delta$). The values
of the resistivity are rather high.  In the pseudogap regime ($\delta
\ll 0.1$) the resistivity tends to turn downward for decreasing
temperature. By contrast, at higher dopings an upward curvature is
observed, leading to a plateau at low $T$, before $\rho$ begins to
drop to lower values at still lower $T$. We cannot exclude that the
plateau is an artifact of the NCA approximation.
\begin{figure}
\begin{center}
\includegraphics[width=0.9 \linewidth]{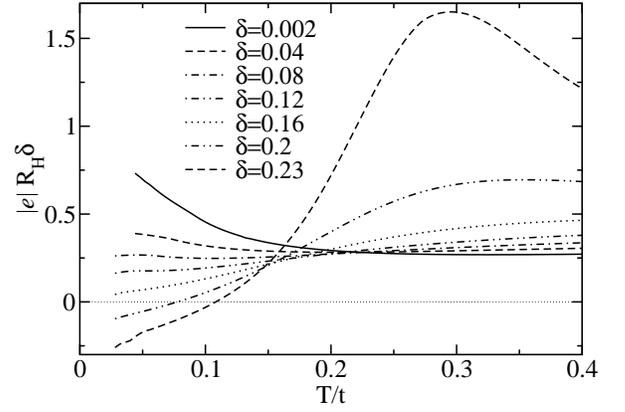}
\end{center}
\caption{\label{RH} $T$-dependence of $R_H$ for $J=0.3 t$. For small
doping and $T\to 0$, $R_H$ approaches the value $1/(|e| \delta)$
expected for a single hole in a $t$-$J$ model.}
\end{figure}
The Hall coefficient is plotted in Fig.~\ref{RH} versus temperature,
for values of $\delta$ ranging from $0.01$ to $0.23$.  For small
doping $\delta < 0.16$, ${\cal R}_H$ is always positive, approaching
the expected value\cite{P_RH} $1/(|e| \delta)$ in the limit
$\delta\to 0, T\to0$.  For doping levels $\delta \geq 0.16$ ${\cal
  R}_H$ is negative at low $T$, consistent with (\ref{48}), and
changes sign at higher $T$, similar to what is observed in experiment
\cite{NTS}.

\section{Instability of the EDMFT solution due to critical
fluctuations in $d=2$}\label{instab.section}

We will now investigate the question of why no solutions of the EDMFT
equations exist for low temperatures and small dopings.  Within the
EDMFT of the $t$-$J$ model long-range antiferromagnetic fluctuations are
not taken into account in a proper way.  As a consequence the local
spin excitation spectrum $\chi_{loc}'' (\omega) = {\rm
  Im}\chi_{loc}(\omega - i0)$ keeps a simple Lorentzian-type shape.
On the other hand the static local susceptibility $\chi_{loc}'(0)$
(in two dimensions) diverges as $\ln\xi$ when the transition to the
antiferromagnetically ordered state is approached and the spin
correlation length $\xi \rightarrow \infty$.  This in turn forces the
slope of $\chi_{loc}''(\omega)$ in the limit $\omega\rightarrow 0$
to diverge as $\ln\xi$ as well.  Within the effective impurity
model of EDMFT a steep slope of $\chi_{loc}''(\omega)$ entails a
large maximum of $\chi_{loc}''(\omega)$ at $\omega_{max}\lesssim
J$, of value $\chi_{loc}''(\omega_{max})\sim \chi_{loc}'(0)$.
As will be shown below, a maximum value of
$\chi_{loc}''(\omega_{max})$ larger than some critical value
$\chi_{loc,crit}'' = \frac{c}{J}$, where the constant $c$ depends
on the density of states $N_J(\epsilon)$ (see (\ref{51})), leads to an
unphysical pole in $\chi_q(\omega)$ at $\omega = \omega_{max}$ and $q
= q_{max}$.  This in turn forces ${\rm Im}M(\omega - i0)$ to change
sign into an unphysical branch of the complex frequency plane.  This
is the point when a stable numerical solution cannot be found any
longer.

To demonstrate this behavior explicitly we consider now a flat density
of states of spin excitations,
\begin{equation}
N_J (\epsilon) = \sum_{\vec q}\delta (\epsilon - J_{\vec q}) =
\frac{1}{8J} \theta (4J - | \epsilon |).
\label{51}
\end{equation}
where the bandwidth $8J$ has been chosen to agree with that of the
tight-binding model, $J_{\vec q} = 2J (\cos q_x + \cos q_y)$.
The local susceptibility as defined by (\ref{8}) may then be expressed
analytically in terms of the self-energy $M(\omega)$
\begin{equation}
\chi_{loc}(\omega - i0) = \frac{1}{8J} \ln \frac{4J + M(\w - i0)}{-4J
+ M(\w - i0)} = \chi' + i\chi''.
\label{52}
\end{equation}
Inverting this relation one finds
\begin{equation}
M(\omega - i0) = 4J \frac{v+1}{v - 1} = 4J \frac{| v|^2 -
1-2iv''}{| v - 1 |^2}\;\;,
\label{53}
\end{equation}
where $v = \exp (8J\chi) = v{'} + iv''$. The imaginary part of $v$,
given by
\begin{equation}
v'' = \exp (8J\chi')\sin (8J\chi'')
\label{54}
\end{equation}
\noindent 
will change sign as $\chi''(\omega)$ increases with increasing
$\omega$, if $8J\chi''\geq \pi$. By (\ref{53}), this will lead to a sign
change of $M''(\omega - i0)$ from negative (stable) to positive
values.  How can $\chi''$ and $M''$ both be positive?  This is
possible since $\chi_q(\omega - i0)$ develops a pole in the physical
domain, $-| J_{\vec q}| < 4J$, at finite $\omega =
\omega_{max}$, giving a contribution to $\chi_{loc}$ with the
``right'' sign.  The instability occurs at finite frequency and thus
is not easily interpreted as a physical phenomenon.  

In the numerical treatment we found that a convergent solution cannot
be obtained when the stability criterion
\begin{equation}
\chi_{loc}'' < \frac{c}{J} 
\label{55}
\end{equation}
\noindent 
is violated.  The constant $c$ takes the value $\pi/8$ for the flat DOS
and a value $\simeq 0.3$ for the tight-binding model.

We emphasize that this instability is not an artifact of the method of
solution of the impurity model but is a generic feature of the EDMFT
equations in two dimensions.  We argue that whenever the ground-state
at $T=0$ is ordered, the self-consistency scheme has to break down
below some finite temperature.  This argument is not only relevant for
our calculation but should be relevant for other applications of EDMFT
which have focused on discussing the possibility of novel quantum
critical points in the presence of two dimensional magnetic
fluctuations\cite{SiNature,SiNumerics}.  While our reasoning does not
apply directly to the quantum critical point, it strongly suggests
that no solution exists on the ordered side of the phase diagram,
casting doubt on the applicability of EDMFT also at the quantum phase
transition.  Our formal argument starts from the observation that in
two dimensions no phase transition (of first or second-order) is
possible for $T > 0$ within EDMFT, since  in a hypothetical ordered phase
the local susceptibility would diverge due to the presence of
Goldstone modes -- in this respect, the EDMFT approach obeys the
Mermin-Wagner theorem.  We mention in passing that even in the case of
Ising symmetry a second-order phase transition is not possible,  as
within EDMFT the longitudinal fluctuations would diverge at the
critical point; however, in this case a first-order transition towards
an ordered phase for $T> 0$ cannot be excluded on general grounds.
Indeed, a first order transition has been found by Sun and
Kotliar\cite{PK} and Zhu, Grempel and Si\cite{GS} for an Ising-coupled
Anderson lattice or Kondo lattice, respectively.  Assuming that for $T
= 0$ the system is magnetically ordered, the local susceptibility
$\chi_{loc} (0)\sim \ln \xi$ will grow steadily as $T$ is lowered
where $\xi$ is exponentially large, $\xi \sim e^{\beta E^*}$, for $T\ll
E^*$ and $E^*$ can crudely be identified with the mean-field
transition temperature. However, we have shown that within EDMFT
$\chi''(\omega)$ is bounded from above by the requirements of
self-consistency.  How can this be reconciled with large
$\chi_{loc}(0) = \int \frac{\text{Im}\chi''(\omega)}{\omega} \frac{d
  \omega}{\pi}\sim \ln \xi$? The only possibility consistent with the
Kramers-Kronig relation is that $\chi''(\omega)$ is constant down to
an {\em exponentially} small energy scale $E^{cr}\sim 1/\xi^z$, where
$z$ is some positive exponent. For sufficiently small $T$, $E^{cr}$
will be exponentially smaller than $T$.  At this point we have to ask
the question whether the solution of the effective impurity model can
result in a scale exponentially smaller than $T$.  We think that this
is extremely unlikely and conclude therefore that no solution can
exist for sufficiently small $T$, consistent with our results and also
with QMC simulations by Burdin {\it et al.} \cite{Burdin} of a model
equivalent to ours in the limit of zero doping.

Eq.~(\ref{53}) also shows how the pseudogap in $M(\omega)$  emerges from the
self-consistency of $\chi$ and $M$.  The absorptive part of the
self-energy $M$, as seen from (\ref{54}), is exponentially small in
the regime where
\begin{equation}
8J\chi{'}(\omega) \gg 1.
\label{56}
\end{equation}
From the numerical results in Fig.~\ref{Mw} one sees that
(\ref{56}) is satisfied if $| \omega| < c_MJ$, where $c_M$ is a
constant of order unity, which depends on the DOS $N_J(\epsilon)$.
Thus, the pseudogap is found to develop as a consequence of the
increase of $\chi{'}(0)\propto \frac{1}{J}\ln\xi$ with growing
$\xi$, in two-dimensions.  We stress that a relation similar to
(\ref{53}) between $M(\omega)$ and $\chi(\omega)$ holds whenever the
DOS $N_J(\epsilon)$ is finite at the band edges, which is a signature
of two dimensions.  In this case the conclusions drawn above
remain valid when $\frac{1}{8J}$ is replaced by $\Delta N_J$, the DOS
jump and $\chi$ is replaced by $\chi - \chi_{reg}$, where
$\chi_{reg}(\omega) = \int d\epsilon (N_J(\epsilon) - \Delta
N_J)/[J_{\vec q} + M(\omega)]$.

\section{Conclusion}

The physics of the doped Mott-Hubbard insulator is governed by the
interplay of the motion of holes and the antiferromagnetic fluctuations
of the spin background.  In this paper we have used a local
approximation scheme to describe both the constrained hopping of holes
and the quantum spin fluctuations in the paramagnetic phase on an
equal footing.  The local approximation becomes exact in the limit of
infinite coordination number of the underlying lattice and is known as
Extended Dynamical Mean Field Theory.  Rather than studying the model
in this limit, we take the point of view that in finite dimensions the
approximation of neglecting the momentum dependence of the single-particle self-energy and the $J$-irreducible spin susceptibility may
still be useful.  Here we have  applied this scheme to the two-dimensional
square lattice with nearest-neighbor hopping and exchange
interaction.  We expect that the approximation should work in a regime
of temperatures and doping concentrations where incoherent
fluctuations dominate and wash out any of the collective effects
sensitive to the system dimension, such as long-range
antiferromagnetic order or superconductivity.

In the regime of temperatures above $T \sim 0.1 J$ and for doping
levels of $0.01 \lesssim \delta \lesssim 0.3$, we
indeed find a highly incoherent phase, with a broad distribution of
spin excitation energies, a high entropy and a large electrical
resistance.  Most strikingly, the local single-particle spectral
function, which is characterized by a narrow peak above the chemical
potential for $\delta \gtrsim 0.25$, develops a pseudogap
as $\delta$ is reduced down to the few percent range. The appearance of
the pseudogap is related to a dramatic shift of the effective chemical
potential from its noninteracting (i.e. Fermi liquid) value near the
center of the lower Hubbard band to the upper band edge.  The shift
persists down to the lowest accessible temperatures of $T \approx 0.1 J$
 and constitutes an unequivocal signal of non-Fermi liquid
behavior in the regime $0.01 \lesssim \delta
\lesssim 0.2$. The single-particle pseudogap is accompanied
by a gap in the spin excitation spectrum for momenta not too close to 
the ordering wave vector $Q = (\pi, \pi)$.

The Hall transport is found to be hole-like, the Hall constant
tending to large positive values $\propto 1/\delta$ as the doping is
reduced.  At large dopings and low temperatures Fermi liquid type
behavior is recovered.

These results are encouraging and give rise to the expectation that
the present EDMFT scheme is able to capture the main features of the
$t$-$J$ model in the incoherent regime.  At lower temperatures and small
dopings one should expect the closeness to the antiferromagnetic
transition at $T=0$ and $\delta < \delta_c$ to play an important
role.  We indeed find that the EDMFT equations stop having a physical
solution below a limiting temperature of $T \approx 0.1 J$.  We are
able to trace this behavior to an intrinsic lack of structure in the
spin structure factor of the effective impurity model, which is
ultimately due to the insufficient treatment of critical fluctuations
in the EDMFT model.  It is likely that similar limitations apply to
other applications of the EDMFT in low dimensional systems.

In conclusion, we emphasize that within the present local
approximation scheme neither effects of finite-range, slowly
fluctuating antiferromagnetic or superconducting domains, nor local
singlet formation or similar short-range correlations are included.
Nonetheless, the strongly incoherent fluctuations characteristic of
our approach (in this case of the spins, but one could imagine similar
effects e.g. in the superconducting sector) suffice to drive
pseudogap formation, a violation of Luttinger's theorem and a
hole-type Fermi surface in the proximity of a Mott insulator.

We acknowledge helpful discussions with E.~Abrahams, J.~Bon\v ca,
A.~Georges, D.R.~Grempel, M.~Grilli, P.~Howell, G.~Kotliar, O.~Parcollet, Q.~Si
and especially P.~Prelov\v sek.  Part of this work was supported by
the Ministry of Education, Science and Sport of Slovenia, the FERLIN
program of the ESF (K.H.) and the Emmy-Noether program of the Deutsche
Forschungsgemeinschaft (A.R.).


\vspace{.5cm}
\appendix
\section{EDMFT derivation}
\label{EDMFT_derivation}

In this appendix we derive the EDMFT self-consistent equations for the
$t$-$J$ model using the cavity method.

To treat the no-double occupancy constraint of the $t$-$J$ model, we will
add a local Coulomb repulsion term explicitly and take the limit
$U\rightarrow\infty$ at the end. In this approach, the electron
creation (destruction) operators $c_i$ ($c_i^{\dagger}$) obey the
usual fermion anticommutation relations. The resulting Hamiltonian is
the 
so-called extended Hubbard model
\begin{equation}
H=-\sum_{ij,\sigma}{t_{ij}c_{i\sigma}^\dagger c_{j\sigma}}+
U\sum_i{n_{i\uparrow}n_{i\downarrow}}+{1\over2}\sum_{ij}J_{ij}\vec{S}_i\cdot\vec{S}_j.
\label{extended_Hubbard_model}
\end{equation}
It is straightforward to extend the theory to other non-local
interactions like non-local Coulomb repulsion, but since we are mainly
interested in the effect of magnetic fluctuations we will neglect
other terms in the Hamiltonian.

For simplicity, let us assume there is no long-range order (i.e. the
system is in the paramagnetic state).  Let us start the derivation of
the EDMFT equations with the action corresponding to the Hamiltonian
(\ref{extended_Hubbard_model}):
\begin{widetext}
\begin{equation}
S =\int_0^{\beta}d\tau \left[
\sum_{ij,\sigma}{c_{i\sigma}^{\dagger}(\tau)\left[\left({\dt}-\mu\right)\delta_{ij}
-t_{ij}\right]c_{j\sigma}(\tau)}
+ { {1\over 2}\sum_{ij}{J_{ij}
\vec{S}_i(\tau)\vec{S}_j(\tau)}+ \sum_{i}{U n_{i\uparrow}(\tau)
n_{i\downarrow}(\tau)}}\right].
\label{extended_Hubbard_action}
\end{equation}
The action can be divided into three parts: the on-site part for the
chosen site ($S_o$)
\begin{equation}
S_o=\int_0^{\beta}{d\tau \Biggl[
\sum_{\sigma}{c_{o\sigma}^{\dagger}(\tau)({\dt}-\mu)c_{o\sigma}(\tau)}+
{U n_{o\uparrow}(\tau) n_{o\downarrow}(\tau)}\Biggr]},
\label{onsite_action}
\end{equation}
the inter-site interaction between the chosen
site $o$ and the rest of the system ($\Delta S$) 
\begin{equation}
\Delta S=
\int_0^{\beta}{d\tau
\Biggl[\sum_{i,\sigma}{-t_{io}c_{i\sigma}^{\dagger}(\tau)c_{o\sigma}(\tau)-
t_{oi}c_{o\sigma}^{\dagger}(\tau)c_{i\sigma}(\tau)} +\Biggr.}
 {\Biggl.{1\over 2}(J_{io}+J_{oi}) \;\vec{S}_i(\tau)\cdot\vec{S}_o(\tau)
\Biggr]},
\end{equation}
and the lattice action in the presence of the cavity ($S^{(0)}$),
which is equal to the original action (\ref{extended_Hubbard_action})
with site $o$ excluded from all summations.

The series expansion in the coupling between the central site and the
rest of the system can be expressed as
\begin{eqnarray}
Z&=&\int Dc_{o\sigma}^{\dagger} Dc_{o\sigma}\int\prod_{i\ne
o} Dc_{i\sigma}^{\dagger} Dc_{i\sigma}
\exp\left(-S_o-S^{(0)}-\int_0^{\beta}\Delta\L(\tau)d\tau\right)\\
&=&
\int Dc_{o\sigma}^{\dagger} Dc_{o\sigma} \exp(-S_o) Z^{(0)}
\left(1-\int_0^{\beta}{\langle \Delta\L(\tau)\rangle^{(0)}}
d\tau+\right.
\left.{1\over 2!}\int_0^{\beta}d\tau_1\int_0^{\beta}d\tau_2{\langle
T_{\tau}\Delta\L(\tau_1)\Delta\L(\tau_2)\rangle^{(0)}}+...\right),
\label{expansion_of_S}
\end{eqnarray}
where $\Delta S=\int_0^{\beta}\Delta\L(\tau)d\tau$ and
$\langle\rangle^{(0)}$ means the average over the cavity action
$S^{(0)}$. In the second line we have integrated out all fermions except
for site $o$.

The first term linear in $\Delta\L$ vanishes, since the average of
each spin $\langle \vec{S}_i(\tau)\rangle=0$ is zero by the assumption
of no long range order in the system. For the broken-symmetry phase,
the spin operator has to be replaced with its deviation from the
average value $\vec{S}_i\rightarrow \vec{S}_i-\langle
\vec{S}_i\rangle$ and  the derivation can proceed along the
same lines.  The second term in the series expansion reads
\begin{multline}
{1\over 2!}\int_0^{\beta}d\tau_1\int_0^{\beta}d\tau_2{\langle
T_{\tau}\Delta\L(\tau_1)\Delta\L(\tau_2)\rangle^{(0)}}= \\
{1\over 2!}\int_0^{\beta}d\tau_1\int_0^{\beta}d\tau_2
\left\langle T_{\tau}
\Biggl[
\sum_{i,\sigma}{t_{io}c_{i\sigma}^{\dagger}(\tau_1)c_{o\sigma}(\tau_1)+
t_{oi}c_{o\sigma}^{\dagger}(\tau_1)c_{i\sigma}(\tau_1)}-
\Biggr.
\right.
\Biggl.
\sum_i J_{oi} \;\vec{S}_o(\tau_1)\cdot\vec{S}_i(\tau_1)
\Biggr]
\\
\times \Biggl[
\sum_{i,\sigma}{t_{io}c_{i\sigma}^{\dagger}(\tau_2)c_{o\sigma}(\tau_2)+
t_{oi}c_{o\sigma}^{\dagger}(\tau_2)c_{i\sigma}(\tau_2)}-
\Biggr.
\left.
\Biggl.
\sum_i J_{io}\;\vec{S}_i(\tau_2)\cdot\vec{S}_o(\tau_2)
\Biggr]
\right\rangle^{(0)}.
\label{no_interference}
\end{multline}
It is crucial to observe that there is no interference between the
kinetic and the spin term since the average of the correlation function
$\left<c_{i\sigma}(\tau_1)\vec{S}_j(\tau_2)\right>^{(0)}$ vanishes.
The leading-order term in the effective action thus reads
\begin{equation}
S_{\text{eff}}=S_o -\iint_0^{\beta}\! d\tau_1 d\tau_2\,\Biggl[
\Biggr.
\Biggl.
{c_{o\sigma}^{\dagger}(\tau_1) \sum_{ij} t_{io}t_{oj}
\left<T_{\tau}c_{i\sigma}(\tau_1)c_{j\sigma}^{\dagger}(\tau_2)\right>^{(0)}
c_{o\sigma}(\tau_2)}+\Biggr.
\Biggl.
{\vec{S}_o(\tau_1)\; {1\over 2}\sum_{ij} J_{io}J_{oj}
\left<T_{\tau}\vec{S}_i(\tau_1)\vec{S}_{j}(\tau_2)\right>^{(0)}\vec{S}_o(\tau_2)}\Biggr]
.
\label{eff_action_e}
\end{equation}
\end{widetext}
Within EDMFT both terms are equally important and are of order 1 in
the $1/d$ expansion.  The two-point Green's function and
the susceptibility scale as $1/d^{|i-j|/2}$ since $t$ and $J$ fall off as
$1/\sqrt{d}$. Furthermore $i$ and $j$ are neighbors of site $o$ and
are thus at least 2 lattice sites apart (in Manhattan distance) giving
a contribution of order $1/d$. The prefactor $t^2$ or $J^2$ is proportional to
$1/d$, while the double sum gives $d^2$ and the net result is
therefore of order 1.

Further it follows from the Linked Cluster Theorem that only {\it
connected} $n$-point correlation functions appear in higher-order
terms of the effective action. Since they have the usual dependence on
$1/d$, all but the first term vanish in the limit
$d\rightarrow\infty$. For instance,the  next-order term would involve
3-point connected correlation function $\chi_{ijk}\sim\left<S_i^z
S_j^z S_k^z\right>$ or $C_{ijk}\sim\left<S_i^z c_j^{\dagger}
c_k\right>$ that scale like $1/d^{|i-j|/2} d^{|i-k|/2}$. When all
three variables $i$, $j$ and $k$ are different, the correlation function
is of order $1/d^2$ since all three sites are neighbors of $o$. The
prefactor $J^3$ or $J t^2$ is proportional to $1/d^{3/2}$ while the  sums
give $d^3$. The term is thus of order $1/\sqrt{d}$. If $i=j$ but
distinct from $k$ the correlation function is of order $1/d$ while
sums give $d^2$ and the net result is again of order
$1/\sqrt{d}$. Higher-order terms fall off faster than
$1/\sqrt{d}$. Thus, in the limit of large $d$ all but the first term
(\ref{eff_action_e}) can be neglected and the effective action becomes
\begin{eqnarray}
S_{\text{eff}}=\int_0^{\beta} U\; n_{o\uparrow}(\tau)n_{o\downarrow}(\tau)
\qquad\qquad\qquad\qquad\qquad
\nonumber\\
-\int_0^{\beta}d\tau_1\int_0^{\beta}d\tau_2\,
{c_{o\sigma}^{\dagger}(\tau_1) {\cal
G}_0^{-1}(\tau_1-\tau_2)c_{o\sigma}(\tau_2)}
\nonumber\\
-{1\over 2}\int_0^{\beta}d\tau_1\int_0^{\beta}d\tau_2\,{\vec{S}_0(\tau_1)
\underline{\chi}_0^{-1}(\tau_1-\tau_2) \vec{S}_0(\tau_2)}
\label{S_eff}
\end{eqnarray}
where
\begin{eqnarray}
{\cal G}_0^{-1}(\iom)&=&\iom+\mu-\sum_{ij}t_{io}t_{oj}G_{ij}^{(0)}(\iom)\;\;,
\nonumber\\
\chi_0^{-1}(\iom)&=&\sum_{ij}J_{io}J_{oj}\;\chi_{ij}^{(0)}(\iom).
\label{Weiss_2}
\end{eqnarray}
The Weiss fields are thus determined by the cavity Green's function $G_{ij}^{(0)}$ and
the cavity susceptibility $\chi_{ij}^{(0)}$.
The absence of interference between the kinetic and spin terms in
(\ref{eff_action_e}) also leads to separate equations for both
cavity quantities
\begin{eqnarray}
G_{ij}^{(0)}&=&G_{ij}-{G_{io} G_{oo}^{-1} G_{oj}}\;\;,\nonumber\\
\chi_{ij}^{(0)}&=&\chi_{ij}-{\chi_{io}\chi_{oo}^{-1}\chi_{oj}}.
\label{Dyson}
\end{eqnarray}
With the power-counting arguments one can show \cite{GGKR,SS} that in
the limit $d\rightarrow\infty$ and EDMFT scaling the single-particle
self-energy $\Sigma(i\omega)$ as well as the double particle
self-energy $M(i\omega)$ become local quantities, i.e.,
\begin{eqnarray}
G_{\vec k}(i\omega) &=& \frac{1}{i\omega + \mu - \epsilon_{\vec k} -
\Sigma (i\omega)}\;\;,\nonumber\\
\chi_{\vec q}(i\omega) &=& \frac{1}{J_{\vec q} + M(i\omega)}.
\label{GFD}
\end{eqnarray}
Inserting the definitions (\ref{GFD}) into (\ref{Dyson}) and
combining with (\ref{Weiss_2}) we finally obtain the
self-consistent conditions
\begin{eqnarray}
{\cal G}_0^{-1} = \Sigma+G_{loc}^{-1}\;\;,\nonumber\\
\chi_0^{-1} = M - \chi_{loc}^{-1}.
\label{SCC}
\end{eqnarray}
These relate the Weiss fields to
the local quantities computable from the local action
(\ref{S_eff}). The system of equations is thus closed.

For practical computation, however, it is convenient to have a
Hamiltonian representation of the local effective action (\ref{S_eff}).
Since it includes retardation effects through frequency
dependent Weiss fields, it is necessary to introduce auxiliary degrees
of freedom describing the baths. The one-particle character of the
Weiss field ${\cal G}_0^{-1}$ can be represented with the fermionic
bath while the two particle field $\chi_0^{-1}$ has a bosonic nature
and dictates bosonic bath. One of the possible choices is
\begin{widetext}
\begin{equation}
H=\sum_{k\sigma}E_k c_{k\sigma}^{\dagger}c_{k\sigma} +
V \sum_{k\sigma}(c_{k\sigma}^{\dagger} c_{o\sigma} +
c_{o\sigma}^{\dagger} c_{k\sigma}) - \sum_{\sigma} {\mu\, c_{o\sigma}^{\dagger}c_{o\sigma}} +
U n_{o\uparrow}n_{0\downarrow}+
\sum_{q}{\omega_q \vec{h}_{q}^{\dagger}\cdot\vec{h}_{\vec
q}}+ I \sum_{q} \vec{S}_o\cdot(\vec{h}_{q}+\vec{h}_{-q}^{\dagger}),
\label{EAM_Ham}
\end{equation}
where $\vec{h}_{q}$ corresponds to a vector-bosonic bath with the
 commutation relations
$[h_{q}^{\alpha},h_{{q}^{\prime}}^{\beta\dagger}]=
\delta_{{q} {q}^{\prime}}\delta_{\alpha\beta}$.
The corresponding action
\begin{multline}
S=S_o+\int_0^{\beta}d\tau\sum_{k\sigma}\left[
c_{k\sigma}^{\dagger}(\tau)
({\dt}+E_k) c_{k\sigma} +
V c_{k\sigma}^{\dagger}(\tau)
c_{o\sigma}(\tau)+V c_{o\sigma}^{\dagger}(\tau)
c_{k\sigma}(\tau)\right]\\
+\int_0^{\beta}d\tau\sum_{q}\left[
\vec{h}_q^{\dagger}(\tau)({\dt}+\omega_q)\vec{h}_q(\tau)+
I\,\vec{h}_q(\tau)\cdot\vec{S}_o(\tau)+{I}\,\vec{S}_o(\tau)\cdot\vec{h}_{-q}^{\dagger}(\tau)
\right]\quad
\end{multline}
is quadratic in $c_{k\sigma}$ and $\vec{h}_q$, and therefore both baths can be
eliminated leading to
\begin{eqnarray}
S=S_o-\int_0^{\beta}d\tau_1\int_0^{\beta}d\tau_2
\sum_{\sigma}c_{o\sigma}^{\dagger}(\tau_1)(\sum_k
V^2{\delta_{\tau_1\tau_2}\over{\dt}+E_k})c_{o\sigma}(\tau_2)-
\int_0^{\beta}d\tau_1\int_0^{\beta}d\tau_2 \vec{S}_o(\tau_1)(\sum_q
I^2 {\delta_{\tau_1\tau_2}\over{\dt}+\omega_q})\vec{S}_o(\tau_2).
\end{eqnarray}
\end{widetext}
This action is identical to the effective action (\ref{S_eff}) provided
that the following relations hold
\begin{eqnarray}
&&{\cal G}_0^{-1}(\tau_1-\tau_2)=-({\dt_1}-\mu)
\delta_{\tau_1\tau_2}+\sum_{k} V^2
{\delta_{\tau_1\tau_2}\over{\dt}+E_k}\;\;,\nonumber\\
&&\chi_0^{-1}(\tau_1-\tau_2)=\sum_q
I^2 \left({\delta_{\tau_1\tau_2}\over{\dt}+\omega_q}+
{\delta_{\tau_1\tau_2}\over-{\dt}+\omega_q}\right)\;\;,
\end{eqnarray}
or equivalently
\begin{eqnarray}
&&{\cal G}_0^{-1}(\iom)=\iom+\mu-V^2 G_{c}(\iom)\;\;, \nonumber\\
&&\chi_0^{-1}(\iom)=-I^2 G_{h}(\iom).
\label{conditions}
\end{eqnarray}
Finally, combining (\ref{SCC}) and (\ref{conditions}) yields
\begin{eqnarray}
G_{loc}^{-1} &=& i\omega+\mu-\Sigma-V^2 G_c\;\;,\nonumber\\
\chi_{loc}^{-1} &=& M+I^2\,G_h\;\;,
\end{eqnarray}
which coincide with (\ref{9}) and (\ref{10}).

\section{Sum rule constraints on $V^2$ and $I^2$}
\label{sumrule}

Within EDMFT, the coupling parameters $V$ and $I$ defined by
(\ref{3old}), describing the hopping on to, and the exchange
interaction with the impurity, are determined self-consistently.
Interestingly, sum rules completely determine the values of $V$ and
partially constrain the values of $I$.  We start with the
single-particle hopping $V$.  Defining a complex variable $z = \omega + \mu +
\Sigma (\omega - i0)$ one may write the EDMFT self-consistency
condition (7), using (2), (8) and (10), as
\begin{equation}
H(z) \equiv \int \frac{D(\epsilon)d\epsilon}{z - \epsilon}
\stackrel{!}{=}\frac{1}{z - V^2 G_c(\omega - i0)}\;\;,
\label{b1}
\end{equation}
with $D(\epsilon)$ the DOS of the tight-binding band $\epsilon_{\vec
k}$ and $G_c(\omega)$ the fermionic bath Green's function  defined in
(12).  Solving (\ref{b1}) for $V^2G_c$ and taking the limit $\omega
\rightarrow \infty$ one finds 
\begin{eqnarray}
\lim_{\omega\to\infty} [\omega V^2G_c(\omega - i0)] = V^2
& = \lim_{\omega\to\infty}\omega (z - \frac{1}{H(z)})\nonumber \\
& = \lim_{\omega\to\infty} \frac{\omega}{z} \langle\epsilon^2\rangle.
\label{b2}
\end{eqnarray}
Here the zero of energy has been chosen such that $\langle\epsilon\rangle = 0$,
with $\langle\epsilon^n\rangle = \int d\epsilon D(\epsilon) \epsilon^n$, and
$\langle\epsilon^2\rangle$ is a measure of the squared width of the band.  For the
tight-binding band $\epsilon_{\vec k} = 2t (\cos k_x + \cos k_y)$ one
finds $\langle\epsilon^2\rangle = 4t^2$.

We now use the sum rule on the spectral weight in the lower Hubbard
band,
\begin{equation}
n_L = \int_{-\infty}^\infty \frac{d\omega}{\pi} {\rm Im} G_{loc}
(\omega - i0) = \frac{1}{2} (1+\delta)
\label{b3}
\end{equation}
which, using the analyticity of
$G_c(\omega - i0)$ in the lower half-plane, is equivalent to the statement
\begin{eqnarray}
n_L &= \lim_{\omega\to\infty} \omega G_{loc}(\omega - i0) =
\lim_{\omega\to\infty}
\frac{1}{N_L}
\sum_{\vec k} \frac{\omega}{z -
\epsilon_{\vec k}}\nonumber \\
&= \lim_{\omega\to\infty} \frac{\omega}{z}\;\;,
\label{b4}
\end{eqnarray}
where $N_L$ is the number of $\vec k$-points in the first Brillouin
zone, which are summed over.
Combining (B2)-(B4) one gets
\begin{equation}
V^2 = \frac{1}{2}(1 + \delta) \langle\epsilon^2\rangle.
\label{b5}
\end{equation}
The coupling constant $I$ may be related to $\chi_{loc}(\omega)$ and
$M(\omega)$.  Using (11), one may express $G_h$ as
\begin{equation}
I^2G_h (\omega) = M(\omega) - \chi_{loc}^{-1} (\omega).
\label{b6}
\end{equation}
Since $G_h$ is a boson Green's function (see (13)) with positive
energy spectrum, $\omega_q \geq 0$, the following relation holds
\begin{equation}
\int_{0}^\infty \frac{d\omega}{\pi} {\rm Im} G_h(\omega
- i0) = 1.
\label{b7}
\end{equation}
From (\ref{b7}) one then finds
\begin{equation}
I^2 = \int_{0}^\infty \frac{d\omega}{\pi}  {\rm Im}
\Big[M(\omega - i0) - \chi_{loc}^{-1}(\omega - i0)\Big].
\label{b8}
\end{equation}
A further relation is obtained by using the f-sum rules:
\begin{equation}
\int \frac{d\omega}{\pi} \omega {\rm Im} G_h (\omega - i0) =
\lim_{\omega\to\infty} \omega^2 G_h (\omega) = \frac{1}{N_L}
\sum_{\vec q} 2\omega_q \equiv \bar\omega_{\vec q}
\label{b9}
\end{equation}
and
\begin{equation}
\int \frac{d\omega}{\pi} \chi_{loc}'' (\omega) =
\lim_{\omega\to\infty}\omega^2\chi_{loc}(\omega).
\label{b10}
\end{equation}
Taking the limit $\lim_{\omega\to\infty}\omega^2 [.....]$ of (\ref{b6})
one finds
\begin{equation}
\bar\omega_{\vec q} = \frac{\langle\epsilon^2\rangle}{2I^2} \int
\frac{d\omega}{\pi} \omega \chi_{loc}'' (\omega)
\label{b11}
\end{equation}
with $\langle\epsilon^2\rangle$ as defined after (\ref{b2}).



\begin{thebibliography}{0}
 

\bibitem{A} P.W. Anderson, {\it The Theory of superconductivity in the
high $T_c$ cuprates}, (Princeton University Press, Princeton, 1997).
\bibitem{SHN} S. Chakravarty, B.I. Halperin, and D.R. Nelson,
Phys. Rev. B {\bf 39}, 2344 (1989).
\bibitem{BBGW} B.B. Beard, R.J. Birgeneau, M. Greven, and U.-J. Wiese,
Phys. Rev. Lett. {\bf 80}, 1742 (1998).
\bibitem{MV} W. Metzner and D. Vollhardt, Phys. Rev. Lett. {\bf 62},
324 (1989).
\bibitem{GGKR} A. Georges, G. Kotliar, W. Krauth, and M.J. Rozenberg,
Rev. Mod. Phys. {\bf 68}, 13 (1996).
\bibitem{SS} Q. Si and J.L. Smith, Phys. Rev. Lett. {\bf 77}, 3391
(1996); J.L. Smith and Q. Si, Phys. Rev. B{\bf 61}, 5184 (2000).
\bibitem{K} H. Kaj\"uter, Ph. D. Thesis, Rutgers University, (1996).
\bibitem{CKW} T.A. Costi, J. Kroha and P. W\"olfle, Phys. Rev. B {\bf
53}, 1850 (1996).
\bibitem{KW} J. Kroha, and P. W\"olfle, in ``Theoretical Methods for
Strongly Correlated Electrons'', D. S\'en\'echal, A.-M. Tremblay,
C. Bourbournais (eds.), CRM Series in Mathematical Physics (Springer,
New York, 2003); cond-mat/0105491.
\bibitem{B} N.E. Bickers, Rev. Mod. Phys. {\bf 59}, 845 (1987).
\bibitem{I} A. Ino {\it et al.,} Phys. Rev. B {\bf 65}, 094504 (2002);
J. C. Campuzano, {\it et al.} Physica B {\bf 259-261}, 517 (1999).
\bibitem{TM} T. Timusk and B. Statt, Phys. Rep. {\bf 62}, 61 (1999).
\bibitem{TL} J.L. Tallon and J.W. Loram, Physica C {\bf 349}, 53
(2001).
\bibitem{CL} J.R. Cooper and J.W. Loram, J. Phys. I France {\bf 6},
2237 (1996); Y. Ando {\it et al.,} cond-mat/0208096.
\bibitem{HRKW} K. Haule, A. Rosch, J. Kroha, and P. W\"olfle, 
Phys. Rev. Lett. {\bf 89}, 236402 (2002).
\bibitem{Maier} Th. A. Maier, Th. Pruschke, and M. Jarrell,
Phys. Rev. B {\bf 66}, 075102 (2002);
 Th.A. Maier, M. Jarrell, A. Macridin, F.-C. Zhang, preprint cond-mat/0208419; 
C. Huscroft, M. Jarrell, Th. Maier, S. Moukouri, and A. N. Tahvildarzadeh, Phys. 
Rev. Lett. {\bf 86}, 139 (2001).
\bibitem{Phillips} T.D. Stanescu, and P. Phillips, cond-mat/0301254 and
cond-mat/0209118.
\bibitem{R} A. Rosch, Phys. Rev. B {\bf 64}, 174407 (2001) and
references therein.
\bibitem{Bu} R. Bulla, A. C. Hewson, Th. Pruschke, J. Phys. Condens. Matter {\bf 10}, (1998) 8365;
R. Bulla, Adv. Solid State Phys. {\bf 40}, 169 (2000).
\bibitem{GS} D.R. Grempel, and Qimiao Si, cond-mat/0207493; J-X Zhu, D.R. Grempel and Qimiao Si, cond-mat/0304033.
\bibitem{PK} P. Sun and G. Kotliar, Phys. Rev. B {\bf 66}, 85120
(2002); P.~Sung and G. Kotliar, preprint cond-mat/0303539.
\bibitem{Barnes} S. E. Barnes, J.~Phys. {\bf F6}, 1375 (1976); S. E. Barnes, J. Phys. {\bf F7}, 2637 (1977).
\bibitem{AA} A. A. Abrikosov, Physics {\bf 2}, 21 (1965).
\bibitem{PG} O. Parcollet and A. Georges, Phys. Rev. B {\bf 59}, 5341
(1999).
\bibitem{SY} S. Sachdev and J. Ye, Phys. Rev. Lett. {\bf 70}, 3339
(1993).
\bibitem{CSKW} T.A. Costi, P. Schmitteckert, J. Kroha and P. W\"olfle,
Phys. Rev. Lett. {\bf 73}, 1275 (1994).
\bibitem{A1} E. Manousakis, Rev. Mod. Phys. {\bf 63}, 1 (1991).
\bibitem{JP} J. Jakli\v{c} and P. Prelov\v{s}ek, Adv. Phys.{\bf
49}, 1 (2000).
\bibitem{LMCAL} J.W. Loram, K.A. Mirza, J.R. Cooper,
N.A. Athanassopoulon, and W.Y. Lian, Prov. of 10th Anniversary HTS
Workshop Houston, World Scientific, 341 (1996).
\bibitem{IFT} M. Imada, A. Fujimori, and Y. Tokura,
Rev. Mod. Phys. {\bf 70}, 1039 (1998).
\bibitem{M-H} E. M\"uller-Hartmann, Z. Phys. B {\bf 74}, 507 (1989).
\bibitem{K1} A. Khurana, Phys. Rev. Lett. {\bf 64}, 1990 (1990).
\bibitem{KY} H. Kohno and K. Yamada, Progr. Theor. Fiz. {\bf 80}, 623
(1988).
\bibitem{VGJ} P. Voruganti, A. Golubetsov, and S. John, Phys. Rev. B
{\bf 45}, 1395 (1992).
\bibitem{LK} E. Lange and G. Kotliar, Phys. Rev. Lett. {\bf 82}, 1317
(1991).
\bibitem{P_RH} P. Prelov\v{s}ek, Phys. Rev. B{\bf 55}, 9219 (1997).
\bibitem{SiNature} Q. Si, S. Rabello, K. Ingersent, and J. L. Smith, Nature,
                 {\bf 413}, 804 (2001. 
\bibitem{SiNumerics} D. R. Grempel and Qimiao Si, preprint
cond-mat/0207493.
\bibitem{Burdin} S. Burdin, M. Grilli, and D. R. Grempel, preprint cond-mat/0206174.
\bibitem{NTS} T. Nishikana, J. Takeda, and M. Sata, J. Phys. Soc. Jpn. {\bf 63}, 1441 (1994).
\end{thebibliography}
\end{document}